%% file: power.tex
\documentclass[longauth, traditabstract]{aa}

\usepackage{amssymb}
\usepackage{amsmath}
\usepackage{graphicx}
\DeclareGraphicsExtensions{.eps, .ps, .eps .gz, ps.gz}
\usepackage[british]{babel} % For British English hyphenation patterns 
\usepackage{placeins}
\usepackage{subfigure}
\usepackage[draft]{hyperref}
\usepackage{color}

\usepackage{natbib}
\bibpunct{(}{)}{;}{a}{}{,}

\def\citejap#1{\citeauthor{#1}\ \citeyear{#1}}

\newcommand{\SSR}{${SSR}$~}
\newcommand{\TSR}{${TSR}$~}

\def\citejap#1{\citeauthor{#1}\ \citeyear{#1}}
\def\SSR{{\rm SSR}}
\def\TSR{{\rm TSR}}
\def\hompc{\smash{\,h\,{\rm Mpc}^{-1}}}
\def\mpcoh{\smash{\,h^{-1}{\rm Mpc}}}

\begin{document}
%%%%%%%%%

\title{The VIMOS Public Extragalactic Redshift Survey (VIPERS).}
\subtitle{The matter density and baryon fraction from the galaxy power spectrum at redshift $0.6<z<1.1$
\thanks{
Based on observations collected at the European Southern Observatory, Cerro Paranal, Chile, using the Very Large Telescope under programmes 182.A-0886 and partly under programme 070.A-9007.
Also based on observations obtained with MegaPrime/MegaCam, a joint project of CFHT and CEA/DAPNIA, at the Canada-France-Hawaii Telescope (CFHT), which is operated by the
National Research Council (NRC) of Canada, the Institut National des Sciences de l'Univers of the Centre National de la Recherche Scientifique (CNRS) of France, and the University of Hawaii. This work is based in part on data products produced at TERAPIX and the Canadian Astronomy Data Centre as part of the Canada-France-Hawaii Telescope Legacy Survey, a collaborative project of NRC and CNRS. The VIPERS web site is \url{http://www.vipers.inaf.it/}. 
}
}

\titlerunning{The VIPERS galaxy power spectrum}
\author{
S.~Rota\inst{\ref{iasf-mi},\ref{brera}}
\and  B.~R.~Granett\inst{\ref{brera},\ref{unimi}}
\and J.~Bel\inst{\ref{cpt}}
\and L.~Guzzo\inst{\ref{brera},\ref{unimi}}
\and  J.~A.~Peacock\inst{\ref{roe}}
\and  M.~J.~Wilson\inst{\ref{roe}}
\and  A.~Pezzotta\inst{\ref{brera},\ref{bicocca}}
% GROUP A1:
\and S.~de la Torre\inst{\ref{lam}}       
\and B.~Garilli\inst{\ref{iasf-mi}}
\and M.~Bolzonella\inst{\ref{oabo}}      
\and M.~Scodeggio\inst{\ref{iasf-mi}} 
% GROUP A2:
\and U.~Abbas\inst{\ref{oa-to}}
\and C.~Adami\inst{\ref{lam}}
\and D.~Bottini\inst{\ref{iasf-mi}}
\and A.~Cappi\inst{\ref{oabo},\ref{nice}}
\and O.~Cucciati\inst{\ref{unibo},\ref{oabo}}       
\and I.~Davidzon\inst{\ref{lam},\ref{oabo}}   
\and P.~Franzetti\inst{\ref{iasf-mi}}   
\and A.~Fritz\inst{\ref{iasf-mi}}       
\and A.~Iovino\inst{\ref{brera}}
\and J.~Krywult\inst{\ref{kielce}} 
\and V.~Le Brun\inst{\ref{lam}}
\and O.~Le F\`evre\inst{\ref{lam}}
\and D.~Maccagni\inst{\ref{iasf-mi}}
\and K.~Ma{\l}ek\inst{\ref{warsaw-nucl}}
\and F.~Marulli\inst{\ref{unibo},\ref{infn-bo},\ref{oabo}} 
\and W.~J.~Percival\inst{\ref{st-andrews}}
\and M.~Polletta\inst{\ref{iasf-mi},\ref{marseille-uni},\ref{toulouse}}
\and A.~Pollo\inst{\ref{warsaw-nucl},\ref{krakow}} 
\and L.~A.~M.~Tasca\inst{\ref{lam}}
\and R.~Tojeiro\inst{\ref{st-andrews}}
\and D.~Vergani\inst{\ref{iasf-bo}}
\and A.~Zanichelli\inst{\ref{ira-bo}}
% GROUP B:
\and S.~Arnouts\inst{\ref{lam}} 
\and E.~Branchini\inst{\ref{roma3},\ref{infn-roma3},\ref{oa-roma}}
\and J.~Coupon\inst{\ref{geneva}}
\and G.~De Lucia\inst{\ref{oats}}
\and O.~Ilbert\inst{\ref{lam}}
\and L.~Moscardini\inst{\ref{unibo},\ref{infn-bo},\ref{oabo}}
\and T.~Moutard\inst{\ref{halifax},\ref{lam}}
}

\offprints{ S. Rota \\ \email{stefano@lambrate.inaf.it} }

\institute{
INAF - Istituto di Astrofisica Spaziale e Fisica Cosmica Milano, via Bassini 15, 20133 Milano, Italy \label{iasf-mi}%1
\and INAF - Osservatorio Astronomico di Brera, Via Brera 28, 20122 Milano
--  via E. Bianchi 46, 23807 Merate, Italy \label{brera}%2
\and  Universit\`{a} degli Studi di Milano, via G. Celoria 16, 20133 Milano, Italy \label{unimi}%3
\and Aix Marseille Univ, Univ Toulon, CNRS, CPT, Marseille, France \label{cpt}%4
\and Institute for Astronomy, University of Edinburgh, Royal
Observatory, Blackford Hill, Edinburgh EH9 3HJ, UK \label{roe}%5
\and Dipartimento di Fisica, Universit\`a di Milano-Bicocca, P.zza della Scienza 3, I-20126 Milano, Italy \label{bicocca}%6
\and Aix Marseille Univ, CNRS, LAM, Laboratoire d'Astrophysique de
Marseille, Marseille, France  \label{lam}%7
\and INAF - Osservatorio Astronomico di Bologna, via Ranzani 1, I-40127, Bologna, Italy \label{oabo} %8
\and INAF - Osservatorio Astrofisico di Torino, 10025 Pino Torinese, Italy \label{oa-to}%9
\and Laboratoire Lagrange, UMR7293, Universit\'e de Nice Sophia Antipolis, CNRS, Observatoire de la C\^ote d’Azur, 06300 Nice, France \label{nice}%10
\and Dipartimento di Fisica e Astronomia - Alma Mater Studiorum Universit\`{a} di Bologna, viale Berti Pichat 6/2, I-40127 Bologna, Italy \label{unibo}%11
\and Institute of Physics, Jan Kochanowski University, ul. Swietokrzyska 15, 25-406 Kielce, Poland \label{kielce}%12
\and National Centre for Nuclear Research, ul. Hoza 69, 00-681 Warszawa, Poland \label{warsaw-nucl}%13
\and INFN, Sezione di Bologna, viale Berti Pichat 6/2, I-40127 Bologna, Italy \label{infn-bo}%14
\and School of Physics and Astronomy, University of St Andrews, St Andrews KY16 9SS, UK \label{st-andrews}%15
\and Aix-Marseille Université, Jardin du Pharo, 58 bd Charles Livon, F-13284 Marseille cedex 7, France \label{marseille-uni} %16
\and IRAP,  9 av. du colonel Roche, BP 44346, F-31028 Toulouse cedex 4, France \label{toulouse}%17 
\and Astronomical Observatory of the Jagiellonian University, Orla 171, 30-001 Cracow, Poland \label{krakow} %18
\and INAF - Istituto di Astrofisica Spaziale e Fisica Cosmica Bologna, via Gobetti 101, I-40129 Bologna, Italy \label{iasf-bo}%19
\and INAF - Istituto di Radioastronomia, via Gobetti 101, I-40129,
Bologna, Italy \label{ira-bo}%20
\and Dipartimento di Matematica e Fisica, Universit\`{a} degli Studi Roma Tre, via della Vasca Navale 84, 00146 Roma, Italy\label{roma3} %21
\and INFN, Sezione di Roma Tre, via della Vasca Navale 84, I-00146 Roma, Italy \label{infn-roma3}%22
\and INAF - Osservatorio Astronomico di Roma, via Frascati 33, I-00040 Monte Porzio Catone (RM), Italy \label{oa-roma}%23
\and Department of Astronomy, University of Geneva, ch. d’Ecogia 16, 1290 Versoix, Switzerland \label{geneva}%24
\and INAF - Osservatorio Astronomico di Trieste, via G. B. Tiepolo 11, 34143 Trieste, Italy \label{oats}%25
\and Department of Astronomy \& Physics, Saint Mary's University, 923 Robie Street, Halifax, Nova Scotia, B3H 3C3, Canada \label{halifax}%26
}

\date{Received --; accepted --}

\abstract
{ 
We use the final catalogue of the VIMOS Public Extragalactic Redshift Survey (VIPERS) to measure the power spectrum of the galaxy distribution at high redshift, presenting results that extend beyond $z=1$ for the first time. We apply a Fast Fourier Transform technique to four independent subvolumes comprising a total of $51,728$ galaxies at $0.6<z<1.1$ (out of the nearly $90,000$ included in the whole survey). We concentrate here on the shape of the direction-averaged power spectrum in redshift space, explaining the level of modelling of redshift-space anisotropies and the anisotropic survey window function that are needed to deduce this in a robust fashion. We then use covariance matrices derived from a large ensemble of mock datasets in order to fit the spectral data. The results are well matched by a standard $\Lambda$CDM model, with density parameter $\Omega_M  h =\smash{0.227^{+0.063}_{-0.050}}$ and baryon fraction $\smash{f_B=\Omega_B/\Omega_M=0.220^{+0.058}_{-0.072}}$. These inferences from the high-$z$ galaxy distribution are consistent with results from local galaxy surveys, and also with the cosmic microwave background. Thus the $\Lambda$CDM model gives a good match to cosmic structure at all redshifts currently accessible to observational study.
}


\maketitle

%----------------------------- sections  -----------------------------
\input{introduction.tex}

\input{section_2.tex}
\input{section_3.tex}

\input{section_4.tex}

\input{section_5.tex}
\input{section_6.tex}

\input{section_7.tex}

%------------------------- acknowledgements -------------------

%\FloatBarrier
\begin{acknowledgements}
We thank J. Dossett for his help with using the CosmoMC routines.	We acknowledge the crucial contribution of the ESO staff for the management of service observations. In particular, we are deeply grateful to M. Hilker for his constant help and support of this programme. Italian participation in VIPERS has been funded by INAF through the PRIN 2008, 2010, and 2014 programmes. LG, BRG, JB, and AP acknowledge support from the European Research Council through grant n.~291521. OLF acknowledges support from the European Research Council through grant n.~268107. JAP acknowledges support of the European Research Council through grant n.~67093. WJP and RT acknowledge financial support from the European Research Council through grant n.~202686. AP, KM, and JK have been supported by the National Science Centre (grants UMO-2012/07/B/ST9/04425 and UMO-2013/09/D/ST9/04030). WJP is also grateful for support from the UK Science and Technology Facilities Council through the grant ST/I001204/1. EB, FM, and LM acknowledge the support from grants ASI-INAF I/023/12/0 and PRIN MIUR 2010-2011. LM also acknowledges financial support from PRIN INAF 2012. SDLT and MP acknowledge the support of the OCEVU Labex (ANR-11-LABX-0060) and the A*MIDEX project (ANR-11-IDEX-0001-02) funded by the ``Investissements d'Avenir" French government programme managed by the ANR and the Programme National Galaxies et Cosmologie (PNCG). Research conducted within the scope of the HECOLS International Associated Laboratory is supported in part by the Polish NCN grant DEC-2013/08/M/ST9/00664.
\end{acknowledgements}

%---------------------------- references  ---------------------------
%\bibliographystyle{mn2e}

\bibliographystyle{aa}
\bibliography{power,biblio_VIPERS_v2}

\end{document}

%% file: introduction.tex
\section{Introduction}
   
Present-day large-scale structures are thought to have
formed by the gravitational amplification of small initial 
density perturbations. The galaxies that define the cosmic 
web are the complicated result of baryonic 
matter falling into dark-matter potential wells after decoupling,
but the overall pattern of inhomogeneity on large scales still
largely reflects the initial conditions. 
If the initial density field, $\delta({\rm \bf x})$, is a
Gaussian process, then its statistical properties are 
completely described by its two-point correlation
function, $\xi(r)$, or by its power spectrum, $P(k)$.  The
shapes of these functions in the linear regime are
directly predicted by theory and depend on the key cosmological 
parameters, especially the total matter density, $\Omega_M$, and the 
baryon fraction, $f_B=\Omega_B/\Omega_M$. 
There is thus a
notable history of using galaxy surveys to probe the
primordial fluctuations and thereby learn about the 
constitution of the Universe.

Any programme for extracting cosmological information from galaxy 
clustering is complicated by several factors. First, 
small-scale density perturbations eventually evolve
in a non-linear fashion requiring more complex modelling techniques
beyond the simple and robust linear-theory predictions.  This entails 
using $N$-body simulations \citep[e.g.][]{springel05} or approximate 
approaches (e.g. \citejap{smith03}; \citejap{bernardeau02}). 
Secondly, we only measure the clustering of luminous tracers; but
the matter and galaxy fields are connected by a complicated bias relation that may be non-linear, stochastic, and non-local
(e.g. \citejap{defw1985}; \citejap{bardeen86}; \citejap{dekel_lahav99}). Thirdly, maps of the large-scale galaxy distribution are built in
redshift space: radial peculiar velocities alter the observed redshift, which introduces a preferred direction into the
otherwise statistically isotropic clustering pattern (\citejap{kaiser87}).

Significant work has been developed over several decades to
overcome these limitations and build galaxy redshift surveys of the
`local' ($z < 0.1$) Universe capable of obtaining 
cosmological constraints.  These include the Sloan Digital Sky Survey
\citep[SDSS: ][]{york00}, in particular through its luminous red
galaxy (LRG) extension \citep{eisenstein05},  and the Two-degree Field Galaxy Redshift
Survey \citep[2dFGRS: ][]{colless03}. 
Direct measurements of the power spectrum have been  
obtained for the 2dFGRS \citep{percival01,cole05}, 
for the SDSS main galaxy sample \citep{pope04,tegmark04,percival07},
and for the LRGs \citep{eisenstein05,huetsi06,tegmark06,percival07,beutler2016}.
These results provide, together with Supernovae Type Ia and Cosmic Microwave Background (CMB) observations,
one of the pillars of the current $\Lambda$CDM cosmological model.

Beyond the lowest-order dependence of the shape of the power spectrum
on the overall matter density, there is also a dependence
on the contribution of massive neutrinos to the energy budget 
\citep[e.g.][and references therein]{xia12} -- albeit below current
sensitivity if the neutrino masses take the lowest values permitted
by oscillation experiments. Beyond this, the baryon fraction
is reflected in the presence of finer-scale modulations of the power 
spectrum: the Baryonic Acoustic Oscillations (BAO), which were
first seen and exploited by the 2dFGRS and SDSS
\citep{percival01,cole05,eisenstein05}. 
BAO measurements at different redshifts now provide one of the best 
probes of the expansion history of the Universe and thus one of the 
key constraints on the properties of the dark energy that is
assumed to drive the accelerated expansion.

Following this path, recent and forthcoming surveys are pushing
to higher redshifts, both through a desire to extend the distance scale \citep{seo07}, 
and also to reduce statistical
errors (since cosmic variance declines as sample volume increases).
Furthermore, high-$z$ perturbations should be in the linear regime on scales smaller than in the local Universe:
$P(k)$ data can then be used up to a larger wave number
$k_{\rm{max}}$, thus extracting more information from the observations. 
The strategy has been in general one of utilising
relatively low-density tracers ($\bar{n}\sim 10^{-4}\,h^{3}{\rm Mpc}^{-3}$)
to minimise telescope time,
exploiting the typical density of fibres achievable with the available
fibre-optic spectrographs.  This has been the case with the Baryon Oscillation Spectroscopic Survey 
\citep[BOSS: ][]{dawson13}, which exploited the SDSS spectrograph
further, extending the concept pioneered with the LRGs
\citep[e.g.][]{alam16}.  Similarly,
the WiggleZ survey further used the long-lived 2dF positioner on the AAT 4-m telescope, to target UV-selected emission-line galaxies \citep{drinkwater10, blake11, blake11b}. 

Covering a large range of redshifts is also of interest through
the ability to study the evolution of large-scale structure.
Most fundamentally, measurements of the growth of fluctuations through redshift-space distortions (RSD) analysis 
allow us to
discriminate between dark energy models and modifications to
Einstein's theory of gravity \citep[e.g.][]{guzzo08,zhang07,samushia13,dossett15}.
Also, understanding the simultaneous evolution of structure and
galaxy biasing teaches us about both
cosmology and galaxy formation, i.e. the complex relationship between
dark and baryonic matter. High-density surveys of
the general galaxy population (such as 2dFGRS and SDSS Main Galaxy Sample at $z\sim 0.1$) 
are essential for this task, both to sample  the density field 
adequately and to provide a representative census of galaxy types.  
This has been the approach of the VIMOS Public Extragalactic Redshift 
Survey (VIPERS), which currently provides the best 
combination of volume and spatial sampling for a
survey going beyond $z=1$.   Its observations were completed recently
(January 2016), collecting a final sample of nearly 90,000 redshifts
with ${0.5 < z < 1.5}$.  With a high-$z$ volume comparable to that of 
the 2dFGRS, VIPERS allows us to perform cosmological investigations at 
$0.6<z<1.1$ with sufficient
control of cosmic variance.  At the same time VIPERS probes the 
clustering properties of a broad range of galaxy classes that may be 
selected by colour, luminosity and stellar mass 
\citep{marulli13,granett15}.  

In this paper we estimate the spherically averaged
redshift-space galaxy power spectrum
$P(k)$ at two different epochs (${0.6 < z < 0.9}$ and ${0.9 < z < 1.1}$).
We discuss in detail the effects of the survey selection function on 
the measured power and how these can be accurately accounted for to 
recover unbiased estimates of cosmological parameters. To this end, we 
make extensive use of mock catalogues to address the impact of 
non-linear evolution, bias and the survey mask. 

This paper is part of the final set of clustering analyses of the full VIPERS survey, and our intention here is to concentrate on the
overall shape of the power spectrum. Redshift-space distortions
affect this measurement, but a
detailed analysis of the clustering anisotropy and its implications
for the growth rate of structures are given in the companion
papers: via correlation functions \citep{pezzotta16}; 
in Fourier space with the additional investigation of `clipping' high-density regions \citep{wilson16}; and in combination with 
galaxy-galaxy lensing \citep{delatorre16}.
In contrast, we focus on the implications of the shape of $P(k)$
for the matter content of the universe,
and how our high-$z$ measurements compare with inferences from
more local studies. 
These Large Scale Structures (LSS) focused papers are accompanied by a number
of further papers that
discuss the evolution of the galaxy population \citep{cucciati16,gargiulo16,haines16}.

The paper is structured as follows: in Section 2, we present
the real and mock VIPERS data; in Section 3, we describe the methodology
for measuring $P(k)$; we discuss our modelling of $P(k)$ in Section 4 
and  present the results of a likelihood analysis of the VIPERS data in
Section 5; in Section 6 we compare our results with analysis performed with previous surveys and summarise our main conclusions in Section 7. Unless explicitly noted otherwise, in our
computations of galaxy distances we adopt a cosmology characterised
by $\Omega_M = 0.30$ and $\Omega_{\Lambda} = 0.70$.

%% file: section_2.tex
%SECTION
\section{VIPERS}
\label{data}
\begin{figure*}
\centerline{\includegraphics[width=160mm,angle=0]{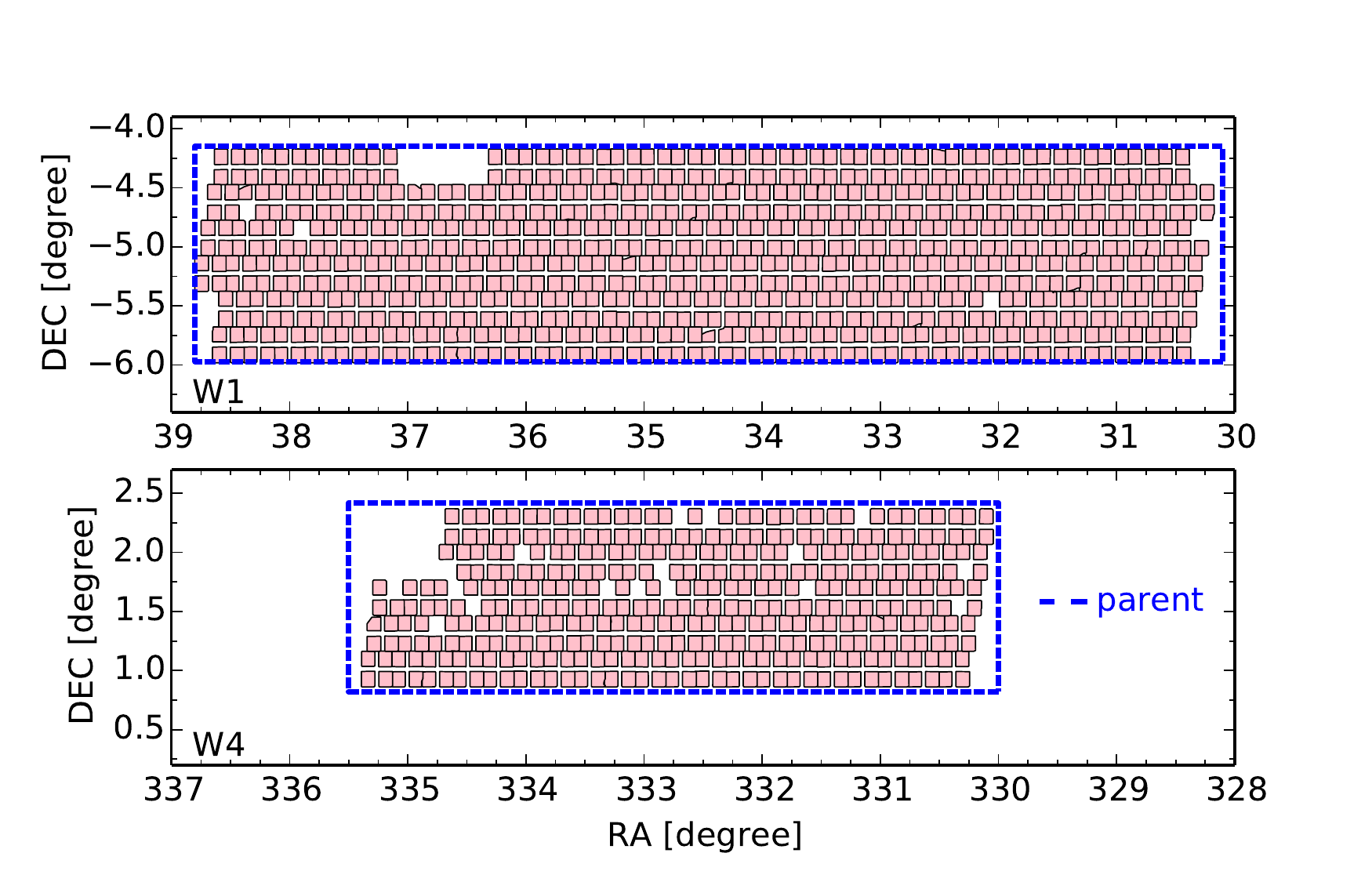}}
\caption{Angular distribution of the full VIPERS galaxy sample, as
used for this study (each pink rectangle corresponds to a single  
quadrant). This shows clearly the geometry and mask produced by the
VIMOS footprint. The dashed blue contours define the area
of what we call the `parent sample'
when we study the survey window function and selection/modelling effects
through the use of mock surveys.}
\label{survey_layout}
\end{figure*}
The VIMOS Public Extragalactic Redshift Survey (VIPERS: \citejap{guzzo14}; \citejap{garilli14}) has used the VIMOS
multi-object spectrograph at the ESO VLT
%\citep[][]{lefevre03} 
to measure redshifts for a sample of almost $90,000$ galaxies with  
$i_{\rm AB} < 22.5$, over a total area of 23.5~deg$^2$.  The VIPERS 
photometric targets were selected from the W1 and W4 fields of the 
Canada-France-Hawaii Telescope Legacy Survey 
Wide\footnote{\url{http://terapix.iap.fr/rubrique.php?id\_rubrique=252}}
\citep[CFHTLS,][]{cuillandre12}, with an additional pre-selection in 
colour that robustly removes galaxies with $z < 0.5$. Because
VIPERS has a single-pass strategy with a maximum target density,
the low-$z$ rejection nearly doubles the
sampling rate at the high redshifts of prime interest.
This yields a mean comoving galaxy density of 
$\bar{n}\sim 5 \times 10^{-3} \, h^{3} \rm{Mpc}^{-3}$ between 
$z=0.6$ and $z\simeq 1.1$.

The survey area was covered homogeneously
with a mosaic of 288 VIMOS pointings (192 in W1,
and 96 in W4), whose overall footprint is displayed in 
Fig.~\ref{survey_layout}.  Spectra were
taken at moderate resolution ($R=220$) using the LR Red grism, 
providing a wavelength coverage of 5500-9500 \AA.  Using a total
exposure time of 45 min, this yields an {\it rms\/} redshift
measurement error (updated using the final PDR-2 data set) which is well described by a relation $\sigma_z=0.00054(1+z)$. 
For further details on the construction and properties of VIPERS see \citet{guzzo14}, \citet{garilli14} and \citet{scodeggio16}. This paper is based on the preliminary version of the Second Public Data Release (PDR-2) sample described in the accompanying paper by \citet{scodeggio16}. The PDR-2 sample includes 340 additional redshifts in the range $0.6 < z< 1.1$ that were validated after this analysis was at an advanced stage. 

\begin{figure}
\centerline{\includegraphics[width=90mm]{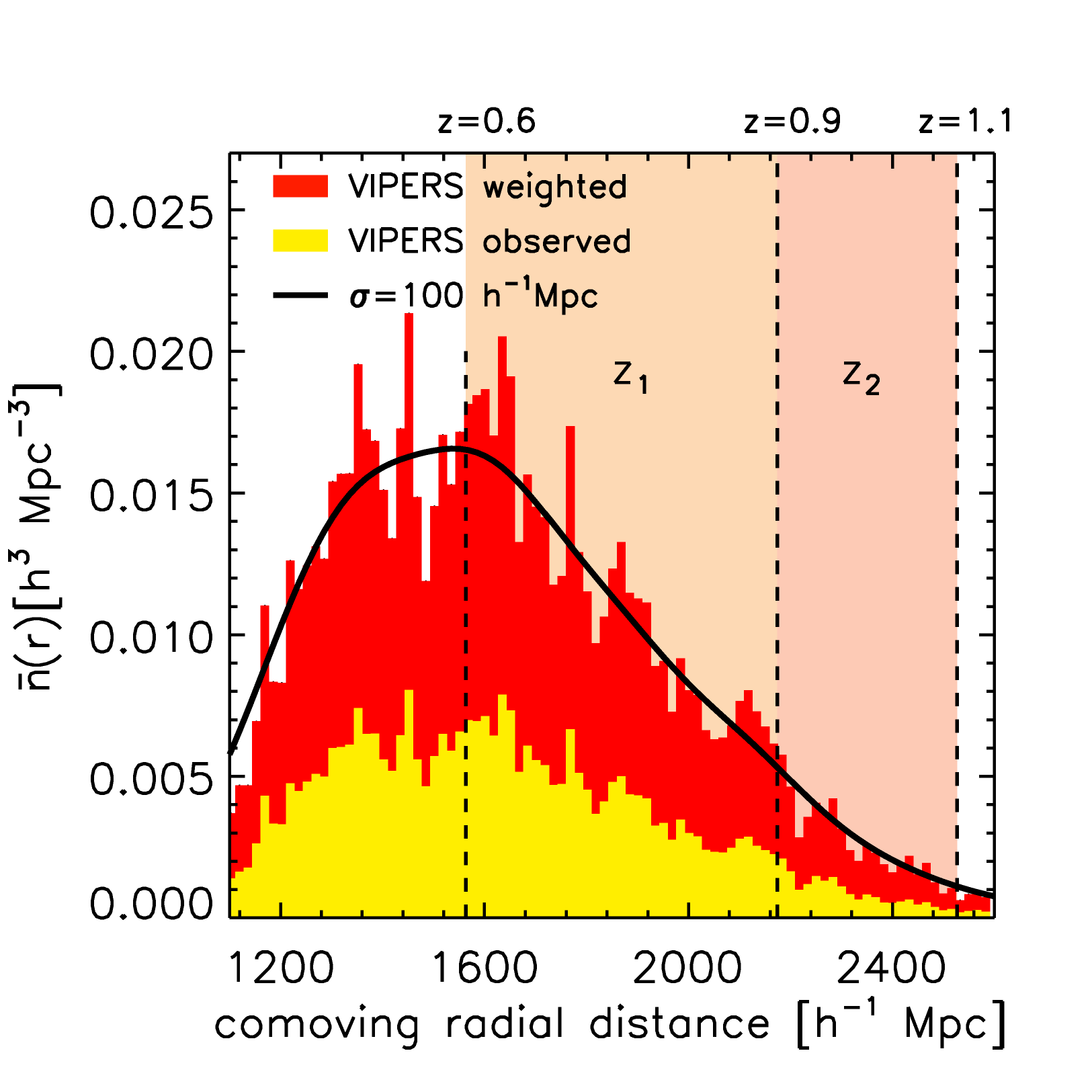}}
\caption{Mean spatial density of galaxies as a function of redshift for the final VIPERS sample used in this analysis. The lower histogram gives the observed distribution, while the top one is obtained after weighting the galaxies with Eq.~\ref{eq:weight}.
The solid line gives the estimated mean density distribution, obtained by Gaussian filtering the observed counts with a kernel of $\sigma=100 \, h^{-1} \rm{Mpc}$.  
The vertical dashed lines define the boundaries of the two redshift ranges analysed here.
}  
\label{real_rd}
\end{figure}

The redshift distribution of the final sample is shown in
Fig.~\ref{real_rd}.  The solid curve shows the corresponding 
distribution expected for an unclustered sample. This is derived from the data by convolving the observed weighted distribution (see next section) with a Gaussian
kernel with standard deviation $\sigma = 100 \mpcoh$. The optimal width of the kernel has been identified by quantifying the impact on the recovered power spectrum from the average of a set of mock samples (see the next section).  We have also compared the results with those using  alternative ways to reconstruct the expected $\left<N(z)\right>$, as used in other VIPERS papers \citep[e.g.][]{delatorre13}, finding no significant differences in the recovered statistics.  

As in all VIPERS statistical measurements, we use only galaxies
with secure redshift measurements, defined as having a quality 
flag between 2 and 9 inclusive and corresponding to an overall redshift confirmation rate of $98\%$ \citep[see][for definitions]{guzzo14}.  

Figure~\ref{real_rd} shows the redshift boundaries of the subsamples defined for this analysis, corresponding to  ${0.6 <  z_1  < 0.9}$ and ${0.9 <  z_2  < 1.1}$. The lower bound at $z=0.6$
fully excludes the transition region produced by the nominal $z=0.5$
colour-colour cut of VIPERS.  In fact, the selection function in this range is well understood, but the gain in volume from adding the $[0.4, 0.6]$ slice would be modest. Conversely, the high-redshift limit at $z=1.1$ excludes the most sparse
distant part of the survey, where shot noise dominates and so
the effective volume is small \citep{tegmark06}.

The mean redshifts for the two redshift samples are $\bar{z}_1 = 0.73$ and 
$\bar{z}_2 = 0.98$. The total numbers of reliable redshifts in
each sample, together with their actual and effective volumes (defined
following \citejap{tegmark06}) are presented in Table \ref{tab}.  
Considering W1 and W4 separately defines four datasets with slightly different window functions. The bias within the two redshift bins will be different due to the different growth
factor and the magnitude-limited nature of the survey.  Owing to the precise photometric calibration of CFHTLS, the target selection is uniform so that we may adopt the same redshift distribution and bias model for both fields.

%FIGURE

%TABLE
\begin{table*}
\centering
\begin{tabular}{|c||c|c|c|c||c|c|c|c|}
\hline
&\multicolumn{4}{|c||}{}&\multicolumn{4}{|c|}{}\\
&\multicolumn{4}{|c||}{W1}&\multicolumn{4}{|c|}{W4}\\ 
&\multicolumn{4}{|c||}{}&\multicolumn{4}{|c|}{}\\ 
\hline
\hline
&&&&&&&& \\
$z$-range & $N_{\rm{gal}}$ & $V$ & $V_{\rm{eff}}$ & $z_{\rm{eff}}$ & $N_{\rm{gal}}$ & $V$ & $V_{\rm{eff}}$ & $z_{\rm{eff}}$ \\
&&&&&&&& \\
\hline
&&&&&&&& \\
0.6-0.9 & 28156 & 9.8 &  9.3 & 0.73  & 14072 & 5.3 & 5.0 & 0.73  \\
&&&&&&&& \\
0.9-1.1 & 6580 & 8.8 & 7.4 &  0.98 & 2920 & 4.7  & 4.0 & 0.97 \\
&&&&&&&& \\
\hline
&&&&&&&& \\
`Full' VIPERS (0.4-1.2) & 48754 &  27.4 &  23.5  & 0.70 & 24323 &  14.8 & 12.7 & 0.70 \\
&&&&&&&&\\
\hline
\end{tabular}
\medskip
\caption{Number of galaxies and volumes for the four subsamples
  analysed in this paper, compared to the `full' survey. Volumes are
  in units of $10^6 \, h^{-3} \rm{Mpc}^3$. Effective volumes are
  defined following 
  \citet{tegmark06}, using a reference galaxy power 
spectrum amplitude of $P_{\rm{eff}}=4000 \, h^{-3} \rm{Mpc}^3$, as obtained at
$k_{\rm{eff}}=0.10 \, h \rm{Mpc}^{-1}$. } 
\label{tab}
\end{table*}

\subsection{Angular masks and incompleteness}

The VIPERS angular selection function accounts for the photometric and
spectroscopic coverage.  Regions around bright stars and with poor
photometric quality in CFHTLS have been excluded giving a loss in area
of 2.5\%.  The spectroscopic coverage is determined by the footprint
of the VIMOS focal plane and the survey strategy, as seen
in Fig.~\ref{survey_layout}.  The spectroscopic
mask results in a survey filling factor of about $\sim 70\%$.
But not all sources in the unmasked area can be
targeted for spectroscopy: the slit assignment algorithm
\citep[SPOC:][]{bottini05} aims to maximise the number of selected
targets with the constraint that spectra may not overlap on the focal plane.  On average 47\% of targets are assigned a slit in the spectrograph. This completeness fraction defines  the target sampling rate (TSR). The result is that close galaxy pairs are missed and the number of spectroscopic targets per quadrant is forced to be approximately constant, independent of the underlying galaxy number density.  The effect is not isotropic on the sky, due to the rectangular shape of the spectral footprint. 
We correct for it as discussed in detail in \citet{pezzotta16}:
we estimate a target sampling rate for each galaxy, $\TSR_i$ within a local region, corresponding to a rectangle with size ${\Delta \rm{RA} \times \Delta \rm{DEC} = 60 \times 100 \, \rm{arcsec^2}}$, slightly larger than the 2D VIMOS spectrum. This is a refinement over the technique used in the PDR-1 VIPERS papers \citep{delatorre13}, in which an average $\TSR$ was used on a quadrant-by-quadrant basis.

Once observed, a target may not produce a reliable redshift
measurement, depending on the galaxy magnitude, the observing
conditions and the available spectral features.  The fraction of
galaxies with reliable redshifts defines the spectroscopic success rate (\SSR), which is $>80$\% over the redshift range analysed here.  The 
$\SSR$ is characterised as a function of galaxy properties and of the VIMOS quadrant, in order to account for varying  observing conditions \citep[see][]{scodeggio16}. 

In addition to the binary photometric/spectroscopic mask, the galaxy selection function is thus given by the product of $\TSR_i$ and $\SSR_i$.  While the binary masks are accounted for by the random sample, this selection function is accounted for in the analysis by weighting each galaxy as 
\begin{equation}
w_i = \frac{1}{\TSR_i \times \SSR_i} \,\,\,\,\, . 
\label{eq:weight}
\end{equation}

We finally note that the colour pre-selection applied to the VIPERS parent photometric sample to isolate galaxies at $z>0.5$ has no effect on this analysis, for two reasons. The Colour Sampling Rate (CSR) \citep{scodeggio16} has in fact been shown to be unity for $z\ge 0.6$; secondly, any residual CSR$(z)$ would not be position-dependent and thus would be absorbed into our model of the redshift distribution.

\subsection{Mock catalogues}
\label{mocks}

To test our algorithms and quantify the level of systematic biases in the final estimate of $P(k)$ and to estimate the expected covariance of our measurements, we used a set of mock galaxy samples built to match the properties of the VIPERS survey.  These are constructed applying a Halo Occupation Distribution (HOD) prescription to dark-matter haloes from a large $N$-body simulation, calibrating the HOD using the actual VIPERS data. The basic procedure described in \citet{delatorre13} and \citet{delatorre16} was applied to generate a new set of mocks based on the Big MultiDark $N$-body simulation (BigMD, \citet{klypin16}). Thanks to its large volume, we were able to generate 306 and 549 mock catalogues for the W1 and W4 fields respectively.  The BigMD assumes 
a Planck-like cosmology with $(\Omega_M,\,\Omega_{\Lambda},\,\Omega_B,\,h,\,n_s,\,\sigma_8\,)$   = (0.307, 0.693, 0.0482, 0.678, 0.960, 0.823).  

We define three types of mock samples to be used in our tests:  (1) the
`parent' mock samples that have the same magnitude and redshift 
limits as the VIPERS sample, but with no angular selection within rectangular 
regions enclosing the full W1 and W4 areas (dashed lines in Fig.~\ref{survey_layout}); 
(2) the `mask' mock samples that exclude galaxies outside 
the angular mask; and (3) the `spectroscopic' mock samples that further 
apply the slit-assignment algorithm in the same manner as the real data.
   

%% file: section_3.tex
%SECTION
\section{Methodology}		
						
\subsection{Power spectrum estimator}
\label{power_spectrum}
	
We estimate the galaxy power spectrum using the method by
\citeauthor{feldman94}
(\citeyear{feldman94}; FKP).  We define the Fourier transform of the density fluctuation field as 
\begin{equation}
\delta(\mathbf{k})=  \int_V \delta(\mathbf{x}) \, \exp^{-i
    \mathbf{k} \cdot \mathbf{x}} \, d^3 \mathbf{x}   \, ,
\end{equation}
where $V$ is the volume of the galaxy sample.  The power spectrum $P(\mathbf{k})$ is
then defined by the variance of the Fourier modes: 
\begin{equation}
\langle \delta(\mathbf{k}) \delta^*(\mathbf{k'}) \rangle = (2\pi)^3
P(\mathbf{k}) \delta_{\rm D}(\mathbf{k} - \mathbf{k'}) \,\,\, . 
\end{equation}
The monopole $P(k)$ is then obtained as the spherical average of $P(\mathbf{k})$ for shells in $k$. 
The practical computation of the  monopole involves binning the data on a Cartesian grid and using a Fast Fourier Transform (FFT) algorithm \citep{jing05,feldman94,frigo02}.
The use of the FFT has also been recently suggested as a way to speed-up the computation of higher order multipole moments of the power spectrum \citep{bianchi15a,scoccimarro15}.  

The FFT decomposes the density field into 
discrete Fourier modes up to the Nyquist frequency
$k_{\rm N}=\pi/H$ with spacing $\Delta k=2 \pi /L$, where $H$ and
$L$ correspond respectively to the distance between two grid points and the total range spanned by the grid.  
Discretising the signal onto
a finite number of cells loses small-scale information leading to aliasing:
small-scale fluctuations beyond the Nyquist frequency become translated to larger scales, creating artefacts in the power spectrum.  These
systematic effects are reduced through the use of a particular mass-assignment scheme (MAS) which applies a low-pass filter.  

A common approach corresponds to convolving the galaxy field with a kernel and then sampling at the positions
of the grid points \citep{hockney88}.  We adopt the Cloud-in-Cell (CIC) as MAS with an explicit expression of the window function in configuration space of $W(\mathbf{x})=\prod\limits_{i=1}^3 W(x_{i}/H)$ , with
\begin{equation}
W(x_{\rm{i}}) = \begin{cases} 1-|x_{i}|, & \mbox{if } |x_{i}| < 1 \\ 0, & \mbox{otherwise} \end{cases}
\label{mas}
\end{equation}

The data are embedded within an FFT cubic
grid with side $ L=800 \, h^{-1}\rm{Mpc}$, and a spacing of $H=2
\,h^{-1}\rm{Mpc}$.  
The corresponding fundamental mode is $k_{\rm{min}}\simeq
0.01\, h  \rm{Mpc}^{-1}$, and hence samples minimum wave number of expected fluctuations along the redshift direction in VIPERS. The smallest scale is
$ k_{\rm{N}}\simeq 1.57\, h 
\rm{Mpc}^{-1}$.  

The normalised density contrast at each point of the FFT grid, $\mathbf{x}_{\rm P}$, is
calculated as
\begin{equation}
\hat{\delta}(\mathbf{x}_{\rm{P}})= \, w(\mathbf{x}_{\rm{P}}) \,
  \frac{n_{\rm{G}}(\mathbf{x}_{\rm{P}})-\alpha \, n_{\rm{R}}(\mathbf{x}_{\rm{P}})}{\sqrt{N}} \, . 
\label{delta}
\end{equation}
Here the $G$ and $R$ labels refer to the galaxy and random
samples.
$n_{\rm R}(\mathbf{x}_{\rm P})$ is the density in a random sample reproducing the full geometry and selection function of the galaxy sample, but with a much higher density than that of
the actual galaxies, $n_{\rm G}(\mathbf{x}_{\rm P})$, so that
the mean inter-particle separation is much smaller than the cell size, $\lambda \ll H$. Outside the survey volume the overdensity is set to 0.
$N$ is defined as
\begin{equation} 
N=\int_{\rm{V}} \bar{n}^2(\mathbf{x}) \, w^2(\mathbf{x}) \, d^3\mathbf{x} =
  \alpha \sum^{N_{\rm R}}_{i=1} \bar{n}(\mathbf{x}_i) \, \tilde w_{\rm{R}}^2(\mathbf{x}_i) \, , 
\label{norm}
\end{equation}
and represents a normalisation factor that accounts for the radial dependence of the mean density in a magnitude-limited survey.   The integral is computed over the total volume of the sample,
$V$, and $\alpha$ is the ratio of the effective total number of galaxies $N_{\rm G}$ to the number of  unclustered random points $N_{\rm R}$:
\begin{equation}
\alpha = \sum_{i=1}^{N_{\rm{G}}} \tilde{w}_{\rm{G}}(\mathbf{x}_i) \; / \; \sum_{j=1}^{N_{\rm{R}}}
\tilde{w}_{\rm{R}}(\mathbf{x}_j).
\end{equation}
In these equations $\tilde{w}_{\rm{G}}(\mathbf{x}_i)$ represents the overall
weight assigned to each galaxy:
\begin{equation}
{\tilde w_{\rm{G}}(\mathbf{x}_i) = w(\mathbf{x}_i) \times w_{\rm FKP}(\mathbf{x}_i) } \,\, ;
\label{eq:weight_total}
\end{equation}
this combines the survey selection function (Eq.~\ref{eq:weight}) with the FKP weight $w_{\rm FKP}(\mathbf{x})$, designed to minimise the
variance of the power  spectrum estimator, under the assumption of Gaussian fluctuations:
\begin{equation}
w_{\rm FKP}(\mathbf{x})=\frac{1}{1+\bar{n}(\mathbf{x})P_{\rm{eff}}(k)} \,\, ,
\label{eq:fkp_weight}
\end{equation}
We choose $P_{\rm{eff}} = 4000 \, (h^{-1} \rm{Mpc})^3$, corresponding to the amplitude of the VIPERS power spectrum  at $k \sim 0.1 \, h \rm{Mpc}^{-1}$.

\citet{percival04b} proposed an extension of the FKP weight to account for the luminosity dependence of bias which in principle can distort the shape of $P(k)$ when estimated from magnitude-limited surveys.  The effect arises since distant sources that are more luminous and have a higher bias contribute most to large-scale modes in the power spectrum.  We have verified that this issue has no detectable effect on the recovered parameters (see Sect.~\ref{consistency_test}).

\noindent Each random point is weighted by 
$\tilde{w}_{\rm R}(\mathbf{x})$ which is equal to the FKP contribution $w_{\rm{FKP}}(\mathbf{x})$.

After Fourier transforming the density field, the monopole 
power spectrum is obtained by averaging in Fourier shells:  
\begin{equation}
\hat{P}(k) = \frac{1}{N_k} \sum_{k < |\mathbf{k'}| < k+ \delta k}  |\hat{\delta}(|\mathbf{k'}|)|^2\;,
\label{power_spectrum_estimator}
\end{equation}
This simple estimator is related to the true power by
aliasing effects arising from the assignment to a mesh
\citep{jing05}:
\begin{equation}	
\begin{split}
\hat{P}(\mathbf{k})=\sum_{\mathbf{n}} |W(\mathbf{k}+2 k_{\rm N}
  \mathbf{n})|^2 \, P(\mathbf{k}+2 k_{\rm N} \mathbf{n}) \\ 
 +  P_{\rm SN}  \sum_{\mathbf{n}} |W(\mathbf{k}+2 k_{\rm N} \mathbf{n})|^2
  \; . 
\label{aliasing}
\end{split}
\end{equation}	
Here, $\mathbf{n}$ is a vector whose components are any integer;
$P_{\rm SN}  $ is the Poisson shot-noise contribution due to the discrete sampling of the density field:
\begin{equation}
\label{shot_noise}
P_{\rm SN} = \frac{\sum_{i=1}^{N_{\rm G}} \tilde{w}_{\rm G}^2(\mathbf{x}_i)+\alpha^2 \sum_{j=1}^{N_{\rm R}} \tilde{w}_{\rm R}^2(\mathbf{x}_j)}{N} \, .
\end{equation}

The importance of choosing $N_{\rm R} \gg N_{\rm G}$ to minimise
contribution to the shot-noise term is evident.

Higher order aliases are damped by the mass assignment window given in Eq.~\ref{mas}; $W(\mathbf{k})$ in Eq.~\ref{aliasing} is its Fourier transform:
\begin{equation}
W(\mathbf{k}) = \left[\prod\limits_{i=1}^3 {\rm sinc} {\left( \frac{\pi k_i}{2 k_{\rm N}} \right)} \right]^{p} .
\end{equation}
For the CIC assignment scheme $p=2$ and in this case the aliasing contribution is 1\% at $k = 0.5  k_{\rm N}$ \citep{Sefusatti16}.  To correct for aliasing requires knowledge of the shape of the power spectrum beyond the Nyquist frequency.   
\citet{jing05} proposed an iterative approach; but the speed of the FFT allows us to push $k_{\rm N}$ to very
high modes by simply reducing the cell size. 
For this reason we choose to correct the estimated 3D power spectrum only
for the first term ($\mathbf{n}=\mathbf{0}$)  
in Eq.~\ref{aliasing}, such that 
\begin{equation}
\hat P(\mathbf{k}) \rightarrow \frac{ \hat{P}(\mathbf{k})-S(\mathbf{k})}{ W({\mathbf{k}})^2} \; .
\label{mas_correct}
\end{equation}
where $S$ is the shot noise contribution.  The aliasing sum arising from the shot noise may be computed analytically in the case when $P_{\rm SN}$ is constant:
\begin{equation}
S(k_1,k_2,k_3) = P_{\rm SN} \times \prod_{i=1}^3  \left[ 1 - \frac{2}{3} \sin^2{\left( \frac{\pi k_i}{2 k_{\rm N}} \right)} \right] \;.
\label{sn}
\end{equation}

\subsection{Survey window function} 
\label{sec:window_function}														

The observed galaxy overdensity field arises by a
multiplication of the true overdensity by the
survey mask: $\delta(\mathbf{x}) \rightarrow \delta(\mathbf{x})\times G(\mathbf{x})$.
In Fourier space this becomes a convolution of Fourier transforms. Provided the two functions have no phase
correlations (fair sample hypothesis), the effect
on the power spectrum is also a convolution
\citep{peacock1991}:
\begin{equation}
P_{\rm{obs}}(\mathbf{k}) = \int P(\mathbf{k}') \, |G(\mathbf{k} - \mathbf{k}')|^2\, \frac{d^3k'}{(2 \pi)^3}\; .
\label{convolution}
\end{equation}
Here, $G(\mathbf{k})$ is the survey window function: the Fourier transform of the mask. We simplify
notation by using the same symbol for the mask and
its Fourier transform; it should always be
clear from context which function is being employed.

In practice, the window must be computed numerically, and
we follow the Monte Carlo approach of
\citet{feldman94}, employing dense random catalogues with the same mask and selection function of the VIPERS data, aligning the redshift direction with the 
$z$ axis.
The number density of random objects, $n_{\rm R}(\mathbf{x})$, is assigned to the grid following the same scheme and each point is weighted using Eq.~\ref{eq:fkp_weight}. The 3D window function in configuration space at each grid-point position is then given by
\begin{equation}
\hat{G}(\mathbf{x}_{\rm P}) = \tilde{w}_{\rm R}(\mathbf{x}_{\rm P}) \, \frac{\alpha \, n_{\rm R}(\mathbf{x}_{\rm P})}{\sqrt{N}} \, ,
\end{equation}
where $N$ is the normalisation factor of Eq.~\ref{norm}. 
After the Fourier transform, the square modulus of $\hat{G}(\mathbf{k})$ is then corrected for the effects of shot noise and the mass-assignment scheme using Eq.~\ref{mas_correct}. 

\begin{figure*}
\centerline{\includegraphics[width=180mm,angle=0]{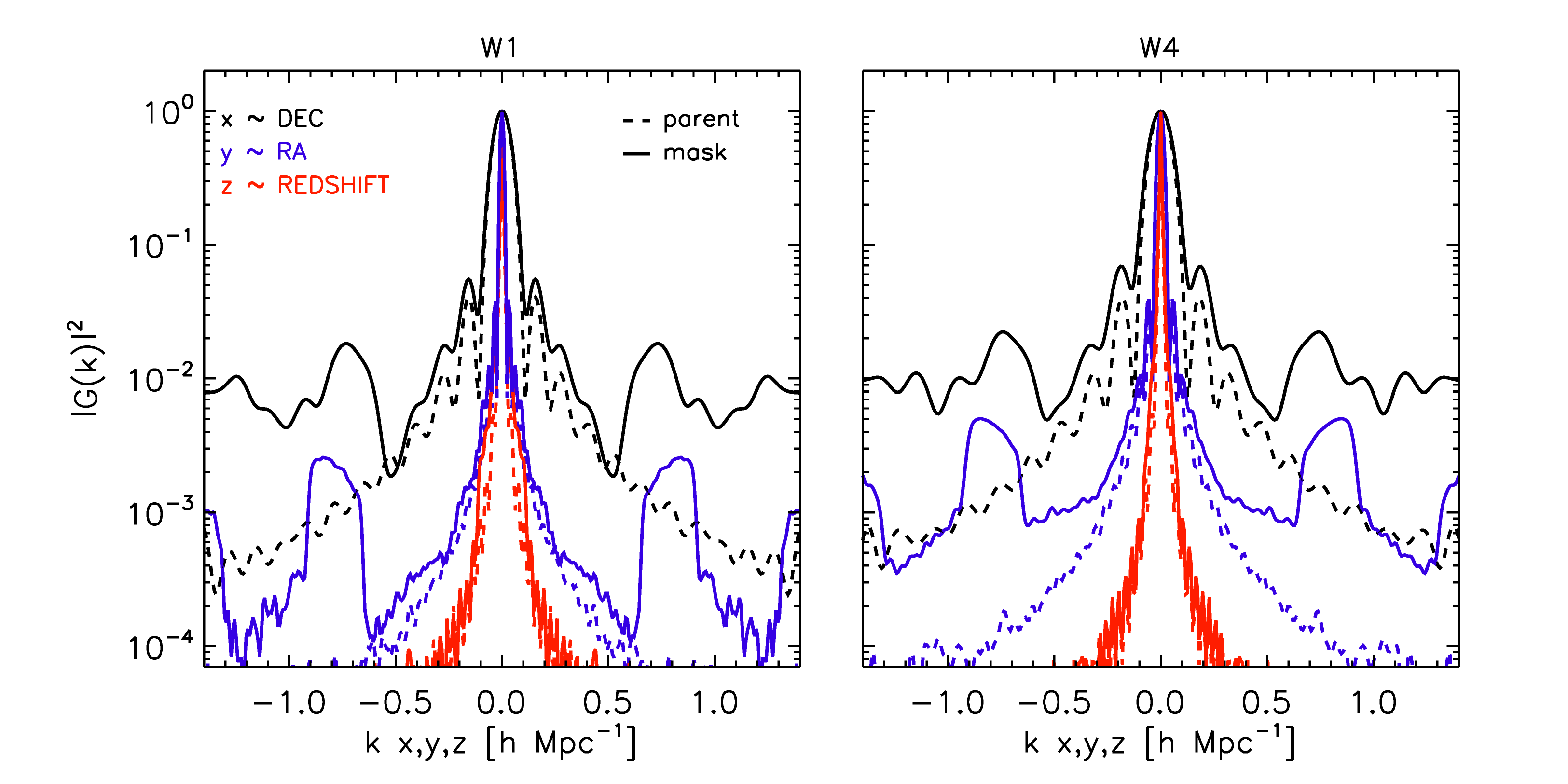}}
\caption{Survey window functions for the W1 and W4 samples at $0.6<z<0.9$, projected along the $k_x$, $k_y$, $k_z$ directions; the dashed lines
  correspond to considering only the geometry of the parent sample,
  while the solid lines give the final window function, when the small-scale angular features of the mask are included. The two samples at $0.9<z<1.1$ show approximately similar window functions.
    }  
\label{window}
\end{figure*}

Figure~\ref{window} shows projections along $k_{x}, k_{y}, k_{z}$ of this estimated window function for the two low-$z$ subsamples.
It is important to note the significantly sharper window function along the redshift direction $k_{\rm{z}}$, compared to the other two axes. It should also be noted how the double extension in right ascension of the W1 field with respect to the W4 one already sharpens the corresponding $|G(k_{\rm{y}})|^2$.  
The effects of the overall geometry (dashed line) and of the survey mask (solid line) are also evident. We note in particular how the small-scale gaps in the VIMOS footprint (see Fig.~\ref{survey_layout}) are reflected in the broad wing features emerging at $k>0.5$. 

\subsection{Accounting for the Window Function}
\label{sec:wf}
In principle, the window could be deconvolved from the measured power spectrum. But the reconstruction can never be perfect, so the errors are complicated to
understand. In contrast, errors of the raw empirical
power spectrum are relatively simple, as discussed by
FKP, and the forward modelling of convolving a
theoretical power spectrum is in principle exact.
We therefore follow this commonly adopted route.

If the power spectrum was isotropic, the 3D integral in Eq.~\ref{convolution}
can be computed over the spherically averaged window.  However, this symmetry is broken in redshift space: owing to the anisotropy introduced by RSD, we must perform the 3D integral first and then spherically average the result. An analytic approximation to obtain the multipoles of $P(\mathbf{k})$ in this case has been proposed by \citet{wilson15}.

The fastest way of calculating the required
convolution is, as usual, to employ the Fast Fourier Transform (FFT).
This means that in fact we transform the
product of the model correlation function and the 
transform of the squared window, where the correlation 
function itself is the transform of the model power:
\begin{equation}	
\begin{split}
P_{\rm{conv}}(\mathbf{k}) = {\rm FFT}^{-1} \, {\rm FFT} \,  \big[G^2 \otimes P_{\rm M} \big] = \\
{\rm FFT}^{-1} \, \big[{\rm FFT} \, G^2 \times 
{\rm FFT} \, P_{\rm M} \big]  \;.
\label{fft_back}
\end{split}
\end{equation}	
Here, $P_{\rm M}$ is the theoretical model and $P_{\rm{conv}}$ the
convolved power spectrum that should be compared with the measured one. This is the approach of \citet{sato10}; its only potential deficiency is that
memory limitations may prevent the FFT mesh
reaching to sufficiently high wave numbers.

Finally, the integral constraint ($IC$) term 
needs to be subtracted from $P_{\rm{conv}}(k)$
to reflect
the fact that the mean density is estimated from the survey volume itself. The power must thus vanish
at $k=0$, requiring
\citep{peacock1991, percival07} 
\begin{equation}	
IC = \frac{|G(\mathbf{k})|^2}{|G(0)|^2} \int   P(\mathbf{k}) \, |G(\mathbf{k})|^2 \, \frac{d^3  k}{(2\pi)^3} \; .
\label{eq42}
\end{equation}	

We test the accuracy of this procedure using our set of mocks.
We run CAMB \citep{lewis11} with the same cosmological parameters of the BigMD, including the 
HALOFIT by \citet{smith03} (which has been updated by \citet{Takahashi2012}) to model non-linearities, to obtain the reference power spectrum
that has to be convolved.  In doing this, galaxy bias is left as a free parameter, an assumption that will be justified for VIPERS in Sect.~\ref{sec:nonlin}.

%FIGURE
\begin{figure*}
\centerline{\includegraphics[width=180mm,angle=0]{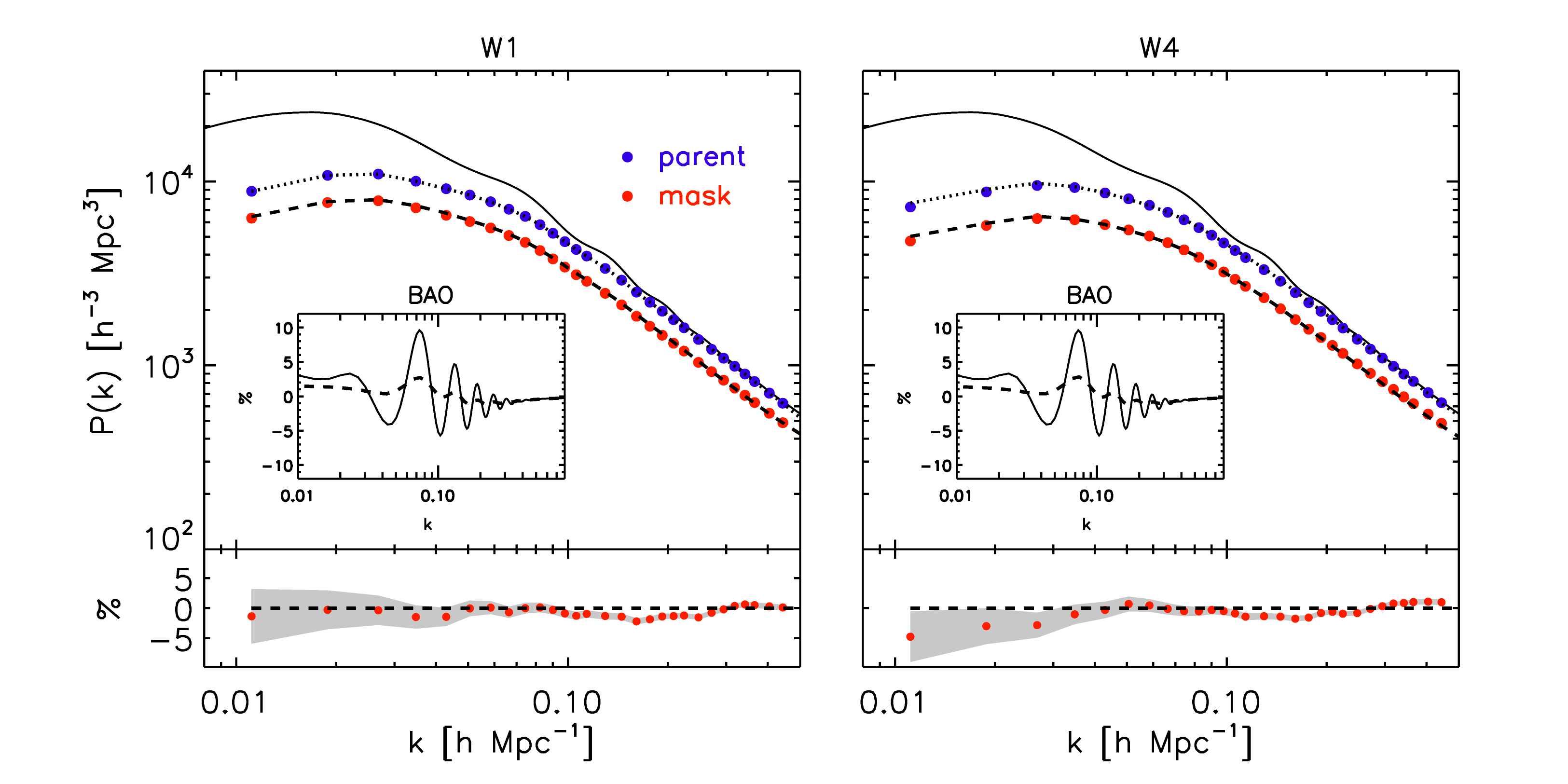}}
\caption{Modelling the effects of the survey window function.  The
  effects of the simplified geometry of the survey only, i.e. the parent
  sample (filled blue circles) and of the full angular mask (filled red
  circles) are compared. The dashed and
  dotted lines show how well these effects are modelled by convolving
  the input $P(k)$ (solid line) with our model for the window functions
  of the two cases. The relative
  accuracy in the case of the full window function (geometry plus
  mask) is explicitly shown in the bottom panel. The insets  show a
  blow-up of the Baryonic Acoustic Oscillations, obtained by
  dividing the input spectrum by a ``no-wiggles" one (solid
  line). This is compared to the actual signal expected when $P(k)$ is convolved with the VIPERS window function (dashed line).} 
\label{bao}
\end{figure*}

Figure~\ref{bao} shows the results of this test, comparing 
the convolved model with the average measurements from the
`observed' mock samples for the two
fields W1 and W4 in the $0.6<z<0.9$ range.  We also distinguish the cases when only the parent survey
geometry is applied to the mocks and when the full angular 
mask  is included. Both of these results are compared with
the model prediction, after
re-scaling for the bias value and convolution with
the appropriate window function. The bottom panels
show residuals, indicating that our modelling can
match the mock results with errors no larger than
1--2\% -- which is as good as perfect for the present
application.

The same figure also shows the dramatic impact of the VIPERS window function on the amplitude of the Baryonic Acoustic Oscillations. 
Nevertheless, as shown in the following section, the overall shape of the power spectrum preserves information on the baryon fraction and other cosmological parameters, once the window function is properly accounted for.

%% file: section_4.tex
%SECTION
\section{Modelling the galaxy power spectrum} 
\label{sec:nonlin}

\subsection{Non-linearity, biasing and redshift-space distortions} 
The measured $P({\bf k})$ is modified by three main effects that need to be taken into 
account in the modelling: (a)
non-linear evolution of clustering, (b) redshift-space distortions and (c) galaxy bias.

In practice, the effect of non-linear
evolution is mitigated by using large-scale data
below a given $k_{\rm{max}}$,
while at the same time adopting an analytical prescription to
account for the residual non-linear deformation of $P(k)$ \citep{smith03,Takahashi2012}.
Over the same range of quasi-linear scales we also assume that galaxy bias can be treated as a simple
scale-independent factor that multiplies
the non-linear matter $P(k)$.  This is consistent with previous studies of the bias scaling and
non-linearity in VIPERS \citep{marulli13,diporto14,cappi15,granett15}.  Moreover the analysis of the mock catalogues in Sect.~\ref{sec:wf} confirms that a constant bias factor is sufficient given that the adopted galaxy formation model is accurate. 

Finally, the measured redshifts are affected by peculiar velocities. 
The present analysis is not concerned with the main
resulting effect, which is
an induced anisotropy of the power spectrum, but
redshift-space distortions will still alter the
spherically averaged power. The linear RSD effect
was analysed by \citet{kaiser87}, who showed that
coherent velocities streaming from underdensities onto
overdensities would introduce a quadrupole anisotropy
in the measured power. This in itself does not change
the shape of the spherically-averaged power: rather,
the amplitude is boosted by a scale-independent factor.
But on non-linear scales, comparable to groups and clusters, galaxies inside virialised structures have random
peculiar velocities. These produce `Finger-of-God' (FoG) radial
smearing that systematically damps modes where the
wave vector runs nearly radially
(e.g. \citejap{peacock94}) -- and this effect 
reduces high-$k$ power even after averaging over directions. 
Thus, an RSD model is required for the present analysis, and 
we employ the simple dispersion model \citep{peacock94}, in which
the \citet{kaiser87} anisotropy is supplemented
by an exponential damping to represent FoG damping:
\begin{equation}
%\rm
P_{\rm s}(\mathbf{k})= b^2 \, P_{\rm r}(\mathbf{k}) \, (1+\beta\mu_{\mathbf{k}}^2)^2 \, e^{-(\sigma_{\rm{TOT}} \, k \, \mu)^2} \, ,
\label{redshift_power}
\end{equation}  
where $P_{\rm s}$ is the redshift-space power spectrum; $\beta=f/b$, where $f$ is the logarithmic growth rate of structure ($f\approx \Omega_M^{\gamma}(a)$, where
$\gamma = 0.55$ for standard gravity); $\sigma_{\rm{TOT}}$ is in units of ($h^{-1} \rm{Mpc}$) and includes the effects of both the velocity dispersion of galaxy pairs $\sigma_v (1+z)/H(z)$ and the VIPERS {\it rms} redshift error (see below); $\mu_{\vec{k}}$ is the cosine of the angle between 
$\mathbf{k}$ and the line-of-sight, which in the FFT grid coincides with the $z$-direction so as to comply with the plane-parallel approximation.

Given the anisotropy of the VIPERS window function (see Fig.~\ref{window}), RSD should be included
in the 3D model $P_{\rm s}(\mathbf{k})$ before convolving with the window as discussed in Sect.~\ref{sec:wf}. This issue was ignored in past
work where the window was more isotropic
(e.g. \citejap{cole05}), but our tests on mock data
show that it is important for VIPERS:
simply convolving the model monopole power spectrum with the 
3D window function yields a poor agreement with the
monopole power taken directly from the mocks.
In contrast, the full 3D modelling method matches the mock
monopole power to a tolerance of just 
a few percent on the scales used in our analysis.

As mentioned above, the effect of the VIPERS redshift measurement errors is considered as an {\it rms} contribution within the RSD Gaussian damping term, as estimated directly from the data: $\sigma_{z} = 0.00054(1 + z)$  or $\sigma_{cz} = 162(1 + z) \, \rm{km \, s^{-1}}$ \citep{scodeggio16}.  $\sigma_{cz}$ is of the order of the dispersion of galaxy peculiar velocity, $\sigma_{v} $.
The two contributions can be added in quadrature, to produce an `effective' pairwise correction to be used in the power spectrum damping factor: $\sigma_{\rm{TOT}} = ( \sigma_{v}^2 + \sigma_{cz}^2 (1+z)^{-2})^{1/2}$. 
This choice has been tested and verified on the mocks.

\begin{figure*}[ht]
\centerline{\includegraphics[width=15cm]{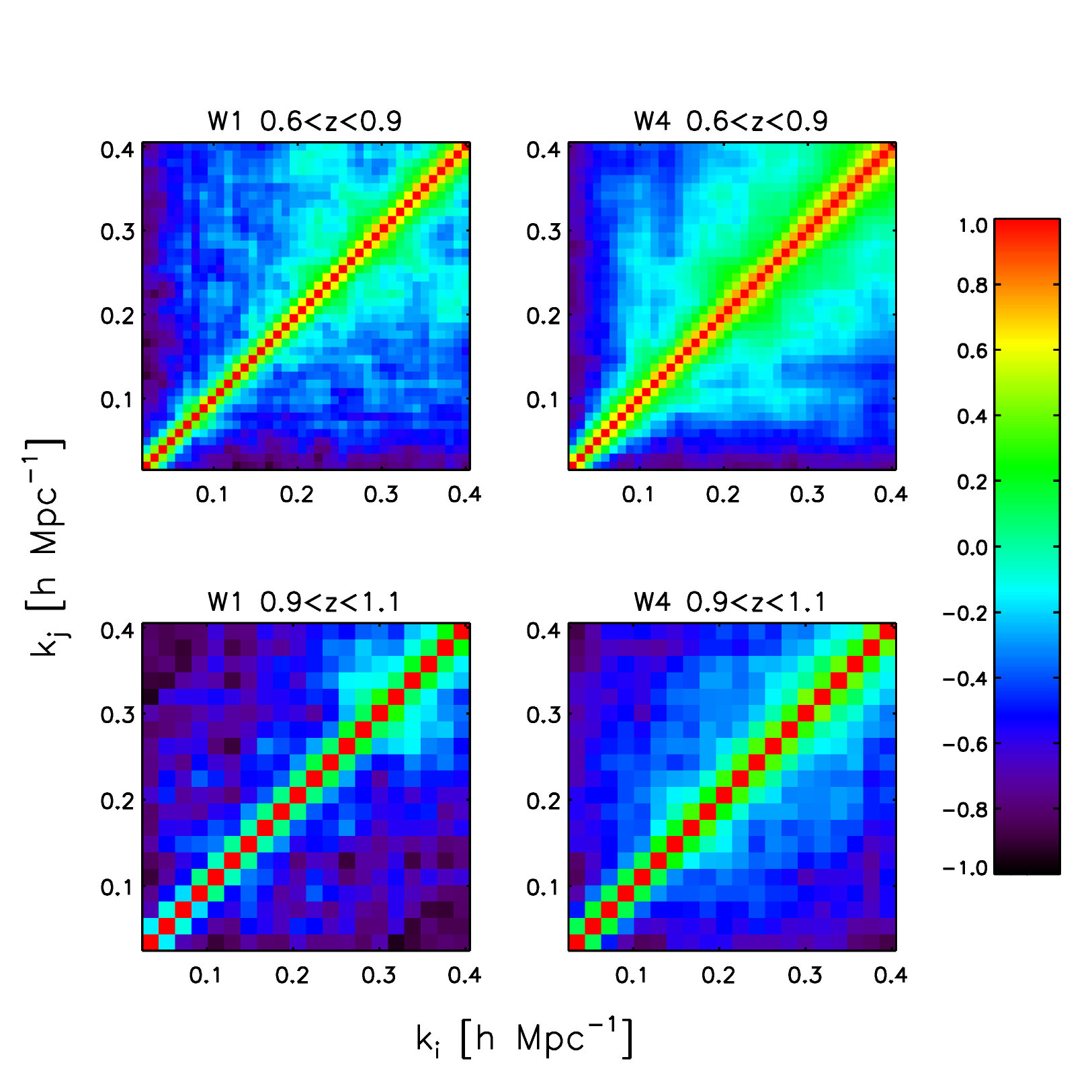}}
\caption{Estimates of the correlation matrices for the four VIPERS subsamples analysed here, constructed using the BigMD mock catalogues as described in Sect.~\ref{cov_section}. It should be noted that the binning is different in the low- and high-redshift samples. Non-Gaussian contributions to the covariance matrix on small scales have greater importance at low redshift.
}  
\label{real_cov}
\end{figure*} 
    
\subsection{Projection effects}

\label{section_fiducial}
The conversion of observed angular coordinates and redshifts into comoving positions introduces an additional dependence on 
the cosmological model \citep{ap1979}. Formally the coordinates must be recomputed for each point in the model parameter space; however, the effort
can be avoided by using the method introduced by \citet{ballinger96} that we follow here.  The following scaling parameters are introduced to express the conservation of the redshift and angular separation of galaxy pairs, $\Delta z$ and $\Delta \theta$: 

\begin{equation}
\alpha = (\alpha_{\perp}^2 \alpha_{\parallel})^{\frac{1}{3}}\;.
\label{corr_fi}
\end{equation}
Here

\begin{equation}
 \alpha_{\perp} = \frac{D_{A,\rm{model}}}{D_{A,\rm{fiducial}}} \; ,
 \end{equation}
where $D_A$ is the angular comoving distance, and

\begin{equation}
 \alpha_{\parallel} = \frac{H_{\rm{fiducial}}( \bar{z} )}{H_{\rm{model}}( \bar{z} )}\; ,
 \end{equation}
where $H( \bar{z} )$ is the Hubble rate at the mean redshift  $\bar{z} $ of the sample.

In the case of a compact survey with fairly isotropic window function, $1/\alpha$ can be used directly to re-scale pair separations $r$ when computing the galaxy correlation function. In the case of $P(k)$, we need to rescale wave numbers as follows:
\begin{equation}
\begin{cases}
k_{\rm{x,fiducial}} & = \; k_{\rm{x,model}} \times \alpha_{\perp} \\
k_{\rm{y,fiducial}} & = \; k_{\rm{y,model}} \times \alpha_{\perp} \\
k_{\rm{z,fiducial}} & = \; k_{\rm{z,model}} \times \alpha_{\parallel}\; ,
\end{cases}
\label{xxx}
\end{equation}
and also multiply the power by $\rm{1/\alpha^3}$. This
must be done before the convolution with the anisotropic window function.

%% file: section_5.tex
%SECTION
\section{Likelihood Analysis} 		
\subsection{Covariance matrix}
\label{cov_section}
We now have in hand all the ingredients needed in order to infer cosmological parameters from the measured clustering power spectrum in VIPERS. However, in order to compute the likelihood of the data for a given model, we need
to know the covariance matrix between the power in different 
modes -- which in general has a complicated non-diagonal structure. This is easily computed if we have a number of
independent realisations of the power spectrum (e.g. from mock data).  The
estimator for the covariance between the $i$  and $j$ power bins is
\begin{equation}
C_{ij} = \frac{1}{N_{\rm{r}}-1} \sum_{m=0}^{N_r} \big[P_{\rm m}(k_{i})-\bar{P}(k_{i})\big]\big[P_{\rm{m}}(k_{j})-\bar{P}(k_{j})\big] \;,
\end{equation}
where $P_{\rm m}(k)$ is one of $N_{\rm r}$ independent estimates of the power spectrum and $\bar{P}(k)$ is the mean.
For a number of bins $N_{\rm{b}}\sim 40$, a few hundred mocks is required in order to obtain a precise covariance matrix \citep{percival14}. The BigMD mocks described in Sect.~\ref{mocks} fulfil this need.  
An unbiased estimate of the inverse covariance matrix is then given by \citep{percival14}
\begin{equation}
\Psi_{ij}  = (1-D) \; C^{-1}_{ij},   \,\, D = \frac{N_{\rm{b}}+1}{N_{\rm{r}}-1}    \,\,\, .
\end{equation}

The covariance is approximately diagonal as would be expected for a Gaussian random field although coupling between modes is evident.  On large scales the dominant effect is due to the window function (Sect.~\ref{sec:wf}).  On small scales the processes of structure formation produce non-Gaussian correlations that are indeed more evident in the low-redshift bin. 

\subsection{Overall Consistency test}
\label{consistency_test}

Before turning to real data, we need to perform an overall test of the modelling pipeline: at which level can our analysis recover an unbiased estimate of the input cosmology of the mock samples, given realistic errors?  

To this end, we have constructed a precise estimate of the monopole power spectrum in two redshift ranges, $0.6 < z < 0.9$ and $0.9 < z < 1.1$, by averaging respectively the $306$ (for W1) and $549$ (for W4) measurements of the corresponding BigMD mocks.
Results are shown here for the low-redshift bin, where non-linearities are expected to be more severe, but we also checked that the same conclusions are valid in the high-redshift sample. 
The two `super-estimates' for W1 and W4 have then been combined in a single likelihood with the models, as we do for the data. Since the volume of W1 is essentially twice that of W4, this is overall equivalent -- in terms of volume -- to a measurement performed over $(306\times 2+549)/3=387$  quasi-independent VIPERS surveys. This leads to an expected reduction of the statistical errors by a factor $\smash{\sqrt{387}}\simeq 20$), but the measurement will be characterised by the same window function and non-linear effects that we have modelled in the previous sections.  

We have obtained theoretical power spectra $P_M(k)$ using CAMB, using the HALOFIT option to give an approximate model of non-linear evolution. The matter density parameter $\Omega_M$, the baryon fraction $f_B$, the Hubble parameter $h$ and the bias $b$ are left free, while all remaining cosmological parameters are fixed to the values used in the BigMD simulation.
Each model power spectrum is derived at our empirical mean redshift, $\bar{z} = 0.73$.   
We then evaluate the VIPERS window function separately for the W1 and W4 fields and perform the 3D convolution using the redshift-space model of Eq.~\ref{redshift_power};
in this expression,
the velocity dispersion $\sigma_v$, with the inclusion of Gaussian redshift errors, 
has been estimated thanks to previous tests where $\Omega_M$ was fixed to the true value. 
Finally, we subtract the integral constraint.

The likelihood between the measurements and the model is then computed accounting for the inverse covariance matrix as estimated above
\begin{equation}
\chi^2(\mathbf{p}) \equiv  \sum_{ij}  \big[P(k_{i})-P_{\rm{M}}(k_{i},\mathbf{p})\big] \, \Psi_{ij} \, \big[P(k_{j})-P_{\rm M}(k_{j},\mathbf{p})\big] \;,
\end{equation}  
where $\mathbf{p}=\{\Omega_M, f_B, h,  b,\sigma_v\}$ is the parameter vector. On four parameters we set flat priors: $\Omega_M$ ($0.2 <  \; \Omega_M < 0.4$),  
$f_B$ ($0 < \; f_B < 0.3$),  
Hubble parameter ($0.6 < \; h < 0.8$), 
bias $(1 < \; b < 2 \;$), 
while for the velocity dispersion we assume a Gaussian prior with a dispersion of $\pm 12 \; \rm{km \, s^{-1}}$ consistent with VIPERS data (see below).  
We consider a restricted range of
wave number, $0.01 < k < k_{\rm{max}}$, estimated in bins spaced by $\Delta k = 0.01\hompc$.  
The minimum value of $k$ corresponds to the maximum extent of the sample in the redshift direction; 
the choice of $k_{\rm{max}}$ has a more critical impact on the precision and accuracy of the parameter estimates. Statistical errors on $P(k)$ are small on smaller scales, but
non-linearities may not be properly modelled if we set
$k_{\rm{max}}$ too high.

\begin{figure}[h!]
\centerline{\includegraphics[width=90mm]{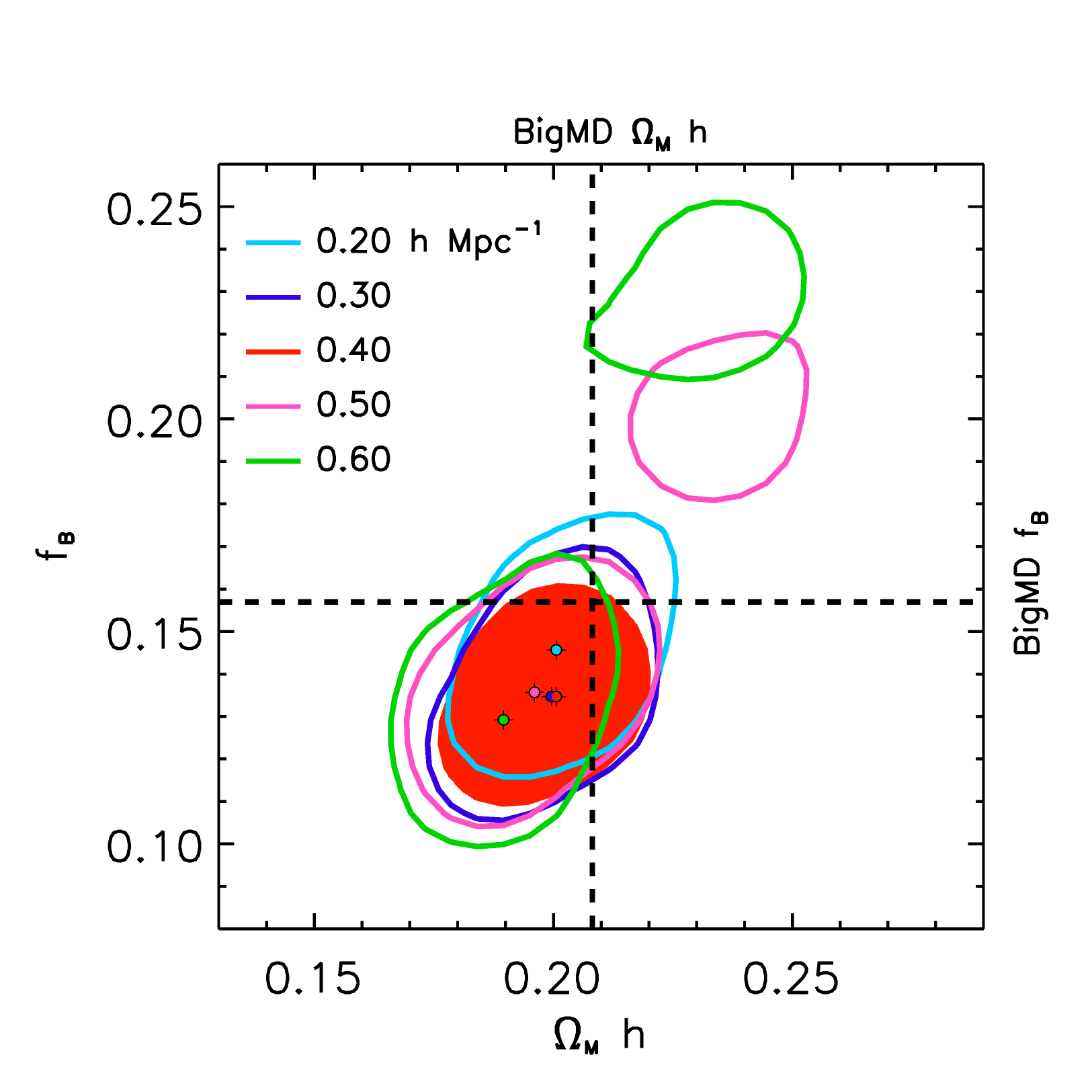}}
\caption{Mean power spectrum from the set of VIPERS mocks used to test the systematic accuracy of the model in recovering the cosmological parameters when including progressively smaller scales. The cosmology of the simulation is indicated by the horizontal and vertical reference lines, and the coloured lines show 68\% confidence levels for different values of $k_{\rm max}$. We find no indication of systematic bias when using scales up to $k_{\rm max}=0.40\hompc$.  Using $k_{\rm max} \ge 0.50\hompc$ we find a degeneracy in the constraints.  We select $k_{\rm max}=0.40\hompc$ for our standard analysis indicated by the filled contour.}

\label{compare_chi}
\end{figure}

In Fig.~\ref{compare_chi}, we test this effect by progressively
increasing $k_{\rm{max}}$ and showing the impact on the
$\chi^2$ contours in the 
$(\Omega_Mh, f_B)$ plane for the simulated combined 
W1 and W4 fields.
We see
that using a maximum wave number $k_{\rm{max}}=0.40 \, h  \rm{Mpc}^{-1}$ (i.e. including $N_{\rm b}=40$ bins in the fit) we are still able to properly describe non-linearities, while excluding significant effects from non-linear bias and slit-exclusion effects.
The values $\Omega_M h \simeq 0.208$  and $f_B \simeq 0.157$ of the BigMD are well recovered even compared to the tiny statistical uncertainty of the `super-mock-sample' used for the test.
In the case of the real VIPERS measurements, the statistical errors will be a factor 
of $\smash{\sim \sqrt{387}}$ larger, indicating that the overall systematic biases in our methodology should be at least an order of magnitude smaller than the error bars.

%% file: section_6.tex
%SECTION 6

%FIGURE
\begin{figure*}
\includegraphics[width=9cm]{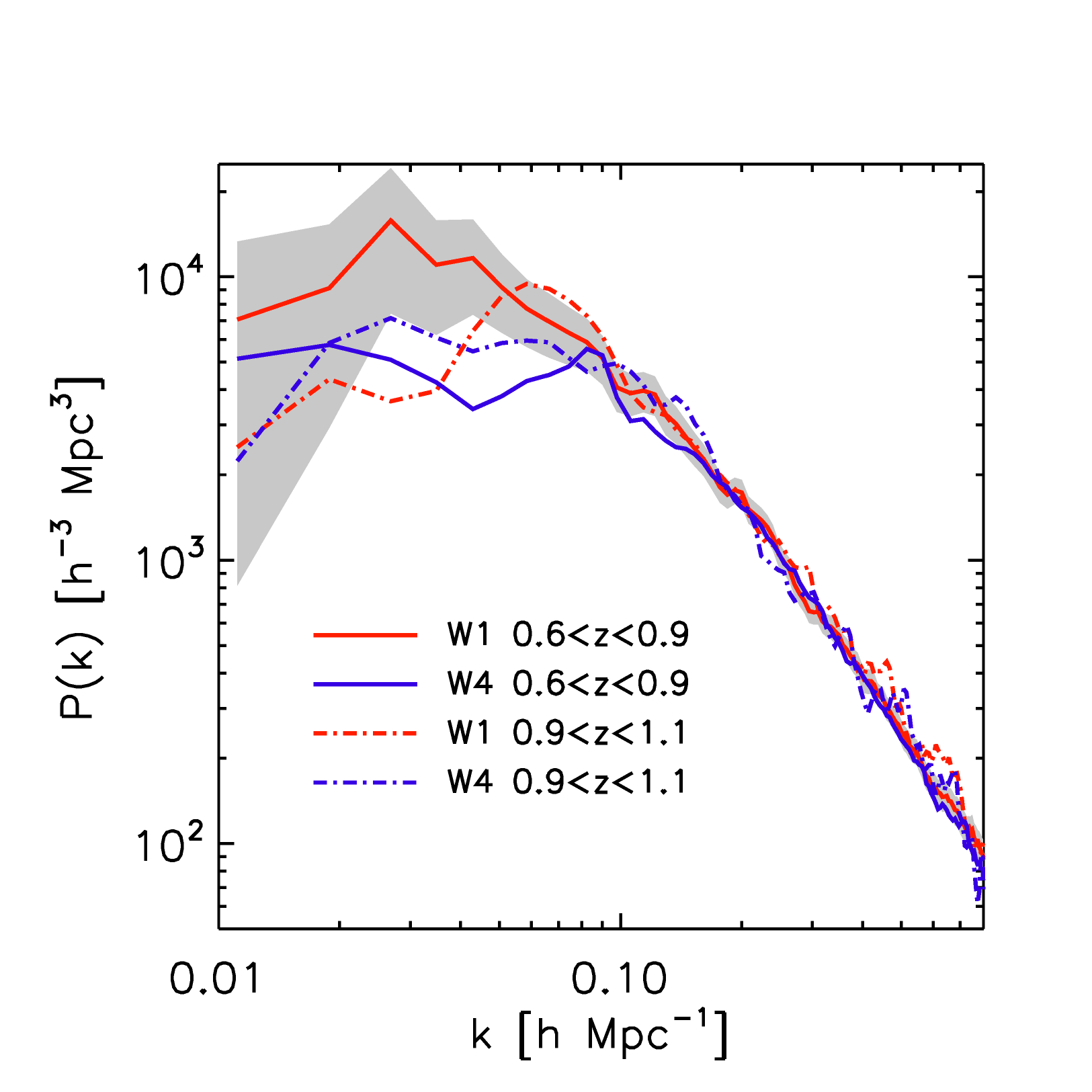}
\includegraphics[width=9cm]{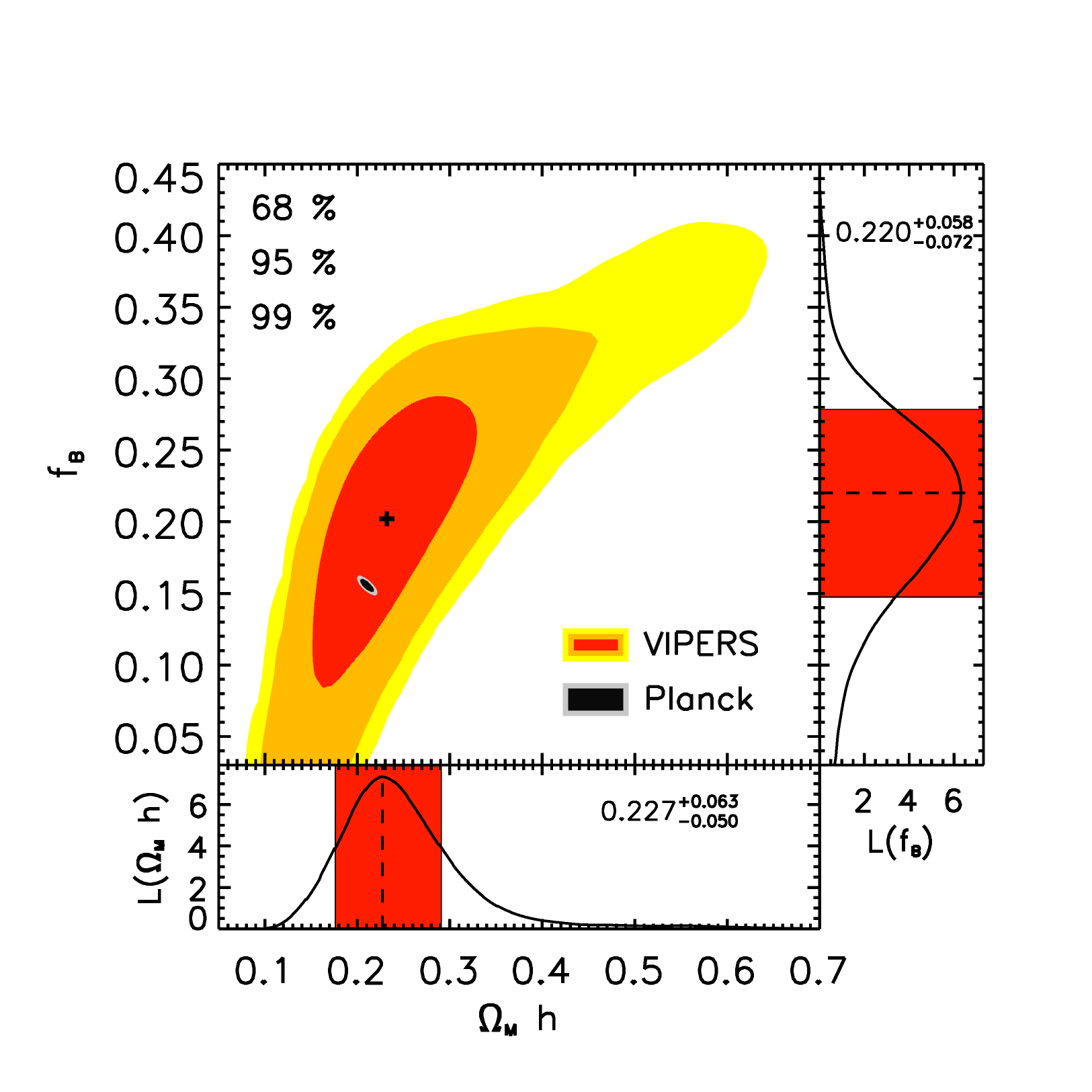}
\caption{{\it Left:} estimates of the monopole of the redshift-space power spectrum from the four independent VIPERS 
subsamples in W1 and W4 and two redshift bins. The shaded area gives the diagonal error corridor around the
$0.6<z<0.9$ W1 sample, as provided by the dispersion of the corresponding mock catalogues. {\it Right:} corresponding likelihood surfaces for the simultaneous fit to the four power
  spectra.   The contours correspond to two-parameter
  confidence levels of  68, 95 and 99 per cent. The measurements have been used down to scales corresponding to $k_{\rm{max}}=0.40\hompc$ and we have marginalised over the galaxy bias and velocity dispersion. }
\label{real_pow}
\end{figure*}

\section{Results}	
\label{sec:result}

\subsection{The VIPERS galaxy power spectrum}

We now apply the machinery developed and tested in 
the previous sections. Figure~\ref{real_pow} 
shows the estimated power spectra from the four subsamples 
of the full VIPERS data defined in Sect.~\ref{data}. The 
grey area indicates the $1-\sigma$ 
error corridor (for one sample only, for clarity; it is 
similar for all samples). Errors correspond to the square root of diagonal elements of the  
covariance matrix, $\sqrt{C_{\rm{ii}}}$.  The contribution of the
shot-noise term is ${P_{\rm SN} \simeq 250 \; h^{-3} \rm{Mpc}^3}$ for the two low
redshift bins and ${P_{\rm SN} \simeq 800 \; h^{-3} \rm{Mpc}^3}$, in the
high-redshift range (due to the sparser galaxy density).  
We have tested with mock samples that the different 
effects of the four window functions on
the overall shape are significant at $k < 0.02\hompc$,
but all four estimates are statistically consistent
in this regime -- thus cosmic variance dominates over any 
differences in the windows. 

The consistency between the high- and low-redshift samples further confirms that the linear biasing and redshfit-space distortion model is adequate over the full redshift range.  This would not necessarily be true if the high-redshift sample suffers from incompleteness which could introduce a scale dependence in the bias.   Any such effect was expected to be small, given the stability of the estimated spectroscopic success rate as a function of redshift and spectral type, at least up to $z=1$ \citep[left panel in Fig.7 of][]{scodeggio16}; nevertheless, the observed consistency in the spectral shape between the two redshift ranges, especially over the scales used in the likelihood analysis, confirms this. Performing the likelihood analysis in the two redshift samples separately, we obtain fully consistent values for the matter density ${\Omega_M h}$ and the 
baryon fraction $f_B$ within the error bars. On the basis of this, we feel even more confident that the power spectra from the two bins can be safely combined into a single likelihood to obtain the VIPERS reference estimates, as we do in the next section.

\subsection{Constraints on the matter density parameter and the baryon fraction}
\label{results}
 
Following \citet{percival01}, \citet{cole05} and \citet{blake2010}, 
we now investigate the constraints that our results can
place on the values of the matter density ${\Omega_M h}$ and the 
baryon fraction $f_B$. 
The density is mainly constrained through the combination 
${\Omega_M}h$, which fixes the wave number corresponding to the
horizon size at matter-radiation equality $k_{\rm{eq}}$,
while the baryon fraction is measured through the amplitude
of the BAO oscillations (the fact that these are small gives very direct evidence that the Universe is dominated
by collisionless matter). Both these aspects are tightly
constrained by the CMB, of course, but for clarity it is
interesting to see what information is given by LSS alone.
However, our analysis cannot be made entirely CMB-free, since
there is a degeneracy with the spectral tilt that is hard to
break. \citet{cole05} showed that the best-fitting value of
$\Omega_M h$ from the 2dFGRS reduced linearly with increasing
$n_s$, with a coefficient of 0.3, and we expect a similar 
coefficient here. We adopt as exact the value
$n_s=0.9677$ \citep{planck2015}, so that $\Omega_M h$ is
raised by 0.010 from the value that would have been obtained
on the assumption of scale-invariant primordial fluctuations.

For other parameters, we adopt relatively broad priors and
checked that our marginalised results are not sensitive
to the prior range.
Unless otherwise noted, we assume a flat $\Lambda$CDM Universe 
with flat priors on $f_B$ $(0   <  f_B  <  0.5)$, $\Omega_M$ $(0.1  <  \Omega_M  <  0.9)$, bias $(1  <   b  <  2)$ and
the Hubble parameter $(0.6 < h < 0.8)$.  The range of bias explored contains the estimates made in previous VIPERS analyses but is large enough to be non-informative, even for the high-redshift bin \citep{marulli13,diporto14,cappi15}.
For the dispersion factor in the
RSD model, we adopt a Gaussian prior for the effective 
velocity dispersion of 
$\sigma_{\rm{TOT}}=257 \pm 12 \, \rm{km \, s^{-1}}$ estimated 
directly from the VIPERS correlation function anisotropy 
\citep{delatorre13,bel14}, which implicitly includes also the 
redshift measurement errors. 

In the analysis of the mock samples we fixed
the normalisation of the power spectrum using the known 
simulation value of ${\sigma_8}$.
Here we fix the scalar amplitude 
${A_s}$ to the best-fit Planck prior (${A_s=2.137 \times 10^{-9}}$),
as this is the quantity directly measured by CMB anisotropy
observations. Our results do not depend strongly on the value of the
scalar amplitude since we marginalise over galaxy bias.

With this set of priors and the machinery for computing the
likelihood as described in Sect. 4 \& 5, we can derive
the posterior likelihood distribution on the parameters of interest.
This is estimated by running MCMC chains on the combined 
W1 and W4 data (accounting for the different window functions), 
while allowing the two
redshift bins to have different bias parameters. Based
on our earlier tests for systematics in analysis of mocks,
we evaluate the likelihood using
the $k$-range $0.01 < k < 0.40\hompc$. The binning size is $\Delta\mathbf{k}=0.01\hompc$ in the low-redshift bin and $\Delta\mathbf{k}=0.02\hompc$ in the high-redshift bin in order to consider the different maximum scale sampled by the two redshift bin volumes.
We thus obtain a marginalised probability density in the
$(\Omega_M h, f_B)$ plane for the whole VIPERS dataset, 
shown on the right in Fig.~\ref{real_pow}. 
The best fit values for the two parameters (after
marginalising over the remaining ones), are, respectively,
$\Omega_M h=0.227^{+0.063}_{-0.050}$ and  $f_B=0.220^{+0.058}_{-0.072}$.
We also obtain marginalised posteriors on the bias values for the two redshift bins, $b=1.32^{+0.14}_{-0.14}$ and $b=1.54^{+0.19}_{-0.14}$, respectively; the increase is consistent with the combination of the intrinsic evolution of bias with the higher mean luminosity of the high-redshift sample.

%
%FIGURE
\begin{figure}
\centerline{\includegraphics[width=8cm]{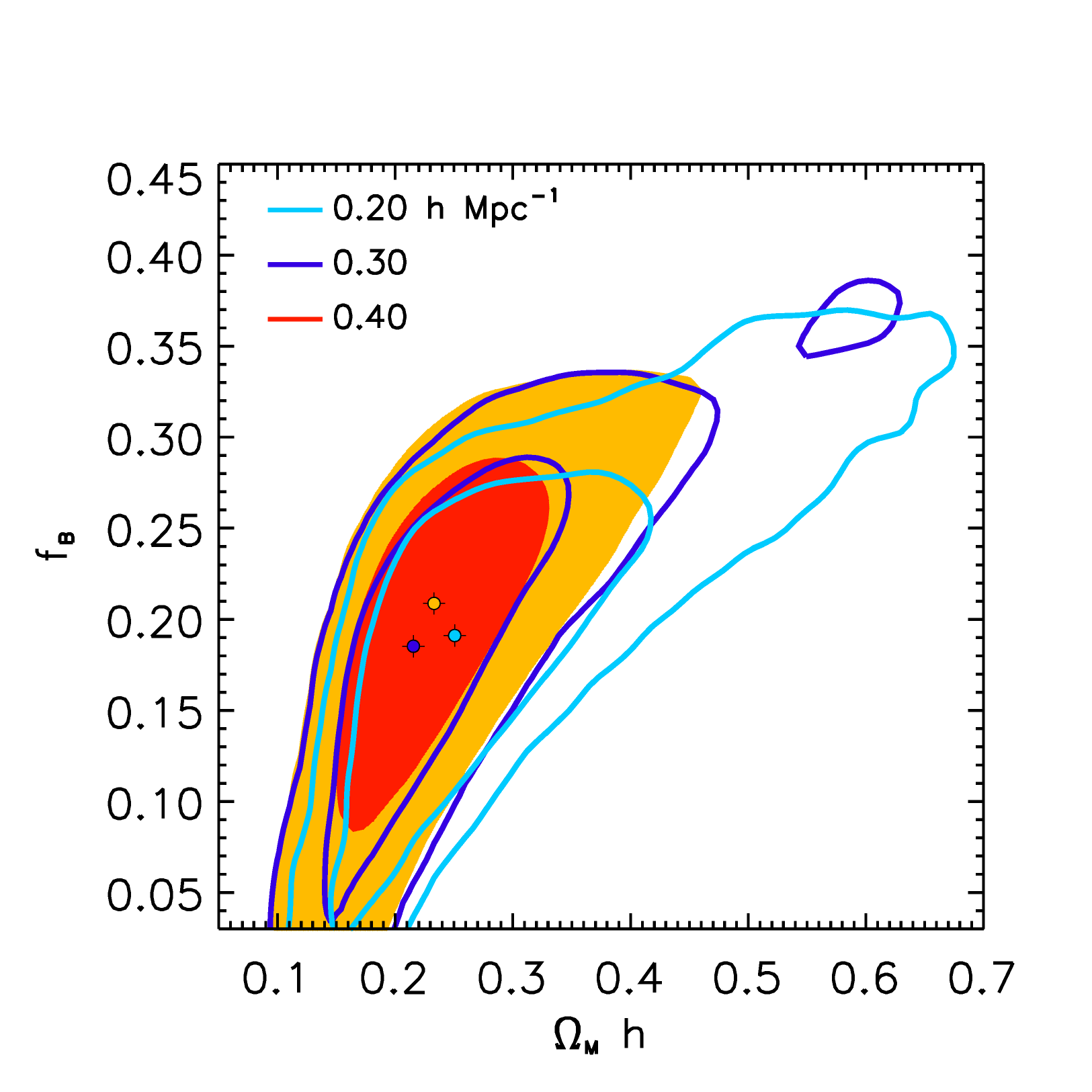}}
\caption{Stability of the estimates of $\Omega_M h$ and $f_B$ when varying the minimum fitting scale $k_{\rm{max}}$. The contours show 68 and 95\% confidence levels.  The filled contour corresponds to $k_{\rm{max}}=0.40\hompc$.  The best-fit from each likelihood analysis is marked with a cross.  No systematic trend with $k_{\rm{max}}$ is evident, confirming the conclusions drawn using the mock samples in Sect.~\ref{consistency_test}.}
\label{real_con}
\end{figure}

\subsection{Stability of the estimates with scale}
\label{maximum_mode}
Our extensive tests with mock data have indicated that our modelling of non-linear effects is fully adequate, with systematic errors well below the statistical uncertainties, which justified extending our likelihood analysis down to $k_{\rm max}=0.40\hompc$. But it is of interest to see whether the results from the real data display the same robustness to variations in $k_{\rm max}$ that we saw in the mocks. 

We have thus repeated our analysis for $k_{\rm max}=0.20$, $0.30$ and $0.40\hompc$.  The results in the plane $(\Omega_M  h, f_B)$ are compared in Fig.~\ref{real_con}. This shows that, as seen with the mock samples, the best-fit values do not change significantly. Naturally, uncertainties
are the largest for the lowest $k_{\rm max}$, since less information (fewer modes) is used.

\subsection{Consistency with VIPERS PDR-1 estimates in configuration space}
\label{pdr1}
Using 60\% of the full VIPERS data (the PDR-1 sample: \citejap{garilli14}), in \citet{bel14} we used the clustering ratio statistic in configuration space \citep[see][for a definition]{bel14a}, to derive the estimate of the matter density $\Omega_M = 0.270^{+0.029}_{-0.025}$.  To carry out a comparison we have repeated our likelihood estimates here with the same priors.  We assume a flat  $\Lambda$CDM cosmological model, described by parameter vector ${\bf p}=\{\Omega_M, \Omega_Bh^2, H_0, A_s, n_s, \sigma_{\rm TOT},b\}$.  A flat prior for the matter density is assumed  ($0.1$--$0.9$), while the other parameters are characterised by Gaussian priors $\Omega_Bh^2=0.0213 \pm 0.0010$  (from BBN: \citejap{pettini08}), $H_0=73.8 \pm 0.024 \,{\rm km\,s^{-1}Mpc^{-1}}$  (from HST: \citejap{riess11}), $\ln(10^{10} A_s)=3.103 \pm 0.072$  and $n_s=0.9616 \pm 0.0094$  \citep[][]{planck16}. 
Bias and effective velocity dispersion $\sigma_{\rm TOT}$ are treated as in Sect.~\ref{results}.

With these prior assumptions the power spectrum data yield a measurement of the matter density parameter 
${\Omega_M=0.261^{+0.027}_{-0.027}}$. 
This value is in excellent agreement with the result of \citet{bel14} but it is
in tension with the 2015 Planck result \citep{planck2015}. 
This apparent discrepancy derives from our adopted prior on $H_{0}$, which differs significantly from the Planck best-fit.
As discussed in \citet{bel14a} it may be reconciled by the fact that the shape of the power spectrum is sensitive to 
the combination $\Omega_M h$ in the linear regime (which becomes $\Omega_M h^2$ on non-linear scales).
As we will see in Sect.~\ref{sec:comparison} the VIPERS constraints on $\Omega_M h$ are 
in much better consistency with the Planck measurements.

%% file: section_7.tex
\section{Discussion and Conclusions}

%
%FIGURE	
\begin{figure*}
\begin{center}
\includegraphics[width=16cm]{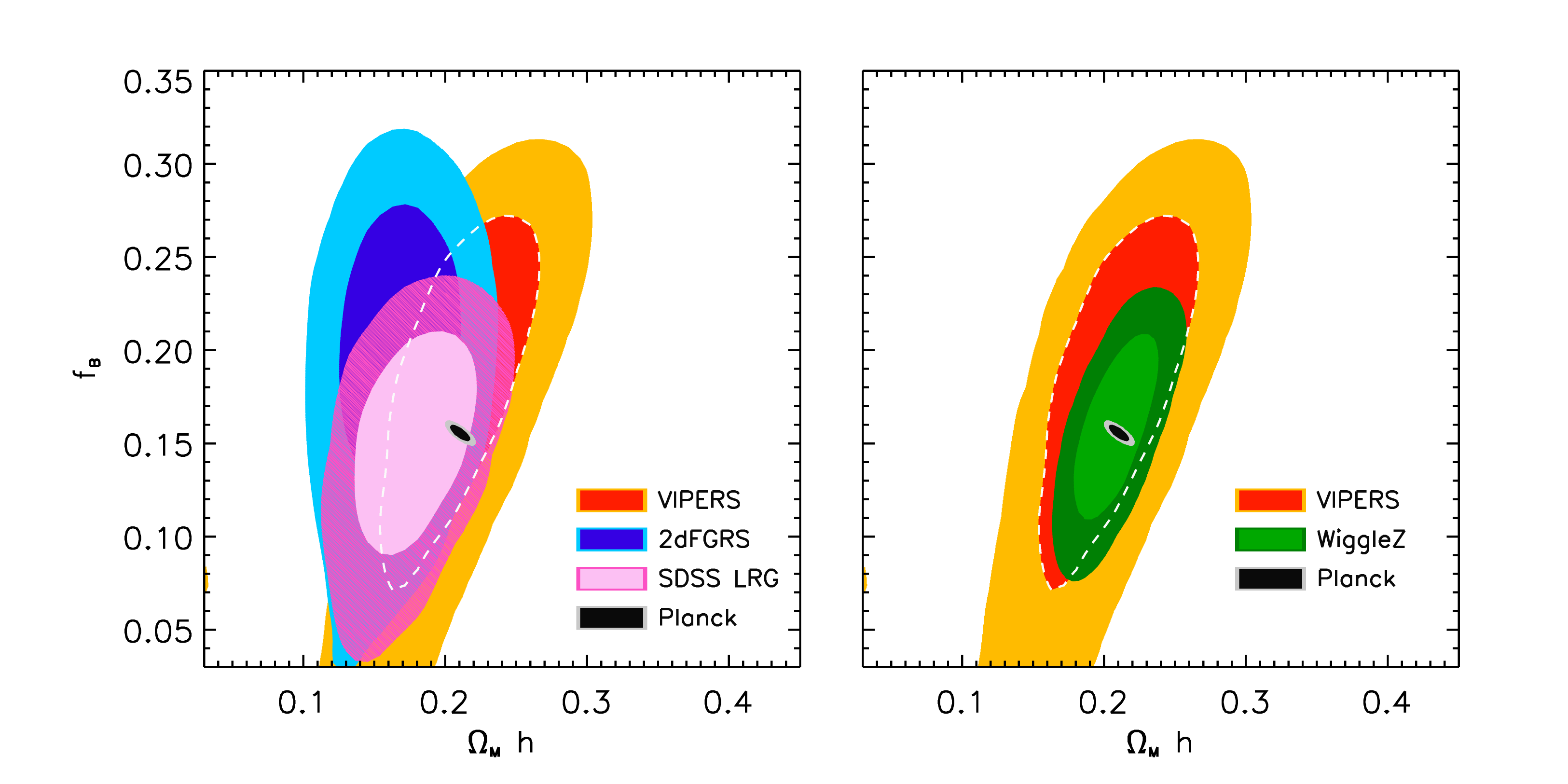}
\end{center}
\caption{Comparison of VIPERS constraints on $f_B$ and $\Omega_M  h$ with other galaxy surveys at low and high redshift. To test consistency with the cosmological model we have fixed the remaining cosmological parameters in the analysis of each survey.
The expansion history is fixed to the fiducial model which restricts the allowed parameter space particularly at high redshift (compare with the VIPERS constraints in Fig. \ref{real_pow}).
  Left panel: at low redshift we show the constraints from 2dFGRS at redshift $z=0.2$ \citep{cole05} and SDSS LRG at redshift $z=0.35$ \citep{tegmark2004}. Right panel: the constraints from WiggleZ \citep{parkinson2012} are shown.    In both panels the constraints from Planck are overplotted \citep{planck2015}.\label{vipers_comparison}}
\end{figure*}

\subsection{Comparison with previous redshift surveys}
\label{sec:comparison}
Comparing the VIPERS measurement with the constraints from datasets at different redshifts provides
a consistency test of the cosmological model.  To perform this test, we analyse other public datasets using the same set of priors as adopted for our own analysis
(Sect.~\ref{results}).

Our prime concern here is to see if the physical shape of the  VIPERS $P(k)$ is consistent with constraints from other galaxy redshift surveys and from the Planck results.
We therefore do not include Alcock-Paczy\'nski effects and choose to
fix the expansion history to the fiducial model with $\Omega_M=0.30$.
This configuration allows us to test the consistency of $P(k)$ shape with maximum statistical power.  We use the
likelihood routines publicly available in the CosmoMC code \citep{cosmomc} to compute the constraints for
2dFGRS \citep{cole05}, SDSS DR4 LRG \citep{tegmark2004} and WiggleZ \citep{parkinson2012} as well as for the Planck 2015 measurements \citep{planck2015}.

The left panel of Fig. \ref{vipers_comparison} shows the constraints on the $(\Omega_M h, f_B)$ plane
from the pioneering 2dFGRS measurements at redshift $z=0.2$ \citep{percival01,cole05}.  The 2dFGRS analysis used power
data up to $k_{\rm max}=0.15\hompc$.  We overplot the constraints from the SDSS LRG sample at redshift $z=0.35$.   The SDSS LRG analysis used power data up to to $k_{\rm max}=0.20\hompc$.
Both analyses marginalise over the parameters of the $Q$-model to fit the scale dependence of the power spectrum on small scales \citep{Cole2005}. We find a lower value of $\Omega_M$ from 2dFGRS, although it is consistent with SDSS and Planck within the 95\% confidence interval. The size of the likelihood region allowed by 2dFGRS and VIPERS is comparable, reflecting  their similar survey volumes. 

The right panel of Fig. \ref{vipers_comparison} provides 
a similar comparison at higher redshift, contrasting the results
of the present paper with
the WiggleZ dataset \citep{parkinson2012,blake2010}.  The WiggleZ analysis used power data up to $k_{\rm max}=0.20 \hompc$ in each redshift bin ranging from $z=0.2$ to $z=0.8$.  This is more conservative than $k_{\rm max}=0.30$ adopted in \citet{parkinson2012}\footnote{\citet{parkinson2012} use a WMAP7 prior which leads to further systematic differences between our results and the constraints shown in Fig. 8 in \citet{parkinson2012}. }.  We find excellent agreement between the VIPERS and WiggleZ constraints and both are consistent with the Planck measurements.

\subsection{Combined constraints}
\label{sec:joint}
Given the consistency of the results found in Sect.~\ref{sec:comparison}, we may combine the constraints on the matter density and baryon fraction from the external LSS surveys.  These constraints are most relevant if we allow the expansion history to vary according to the model; thus we now adopt the methodology described in Sect.~\ref{section_fiducial} to account for the distance scaling. Again, we use the priors described in Sect.~\ref{results}.

\begin{figure}
\begin{center}
\includegraphics[width=8cm]{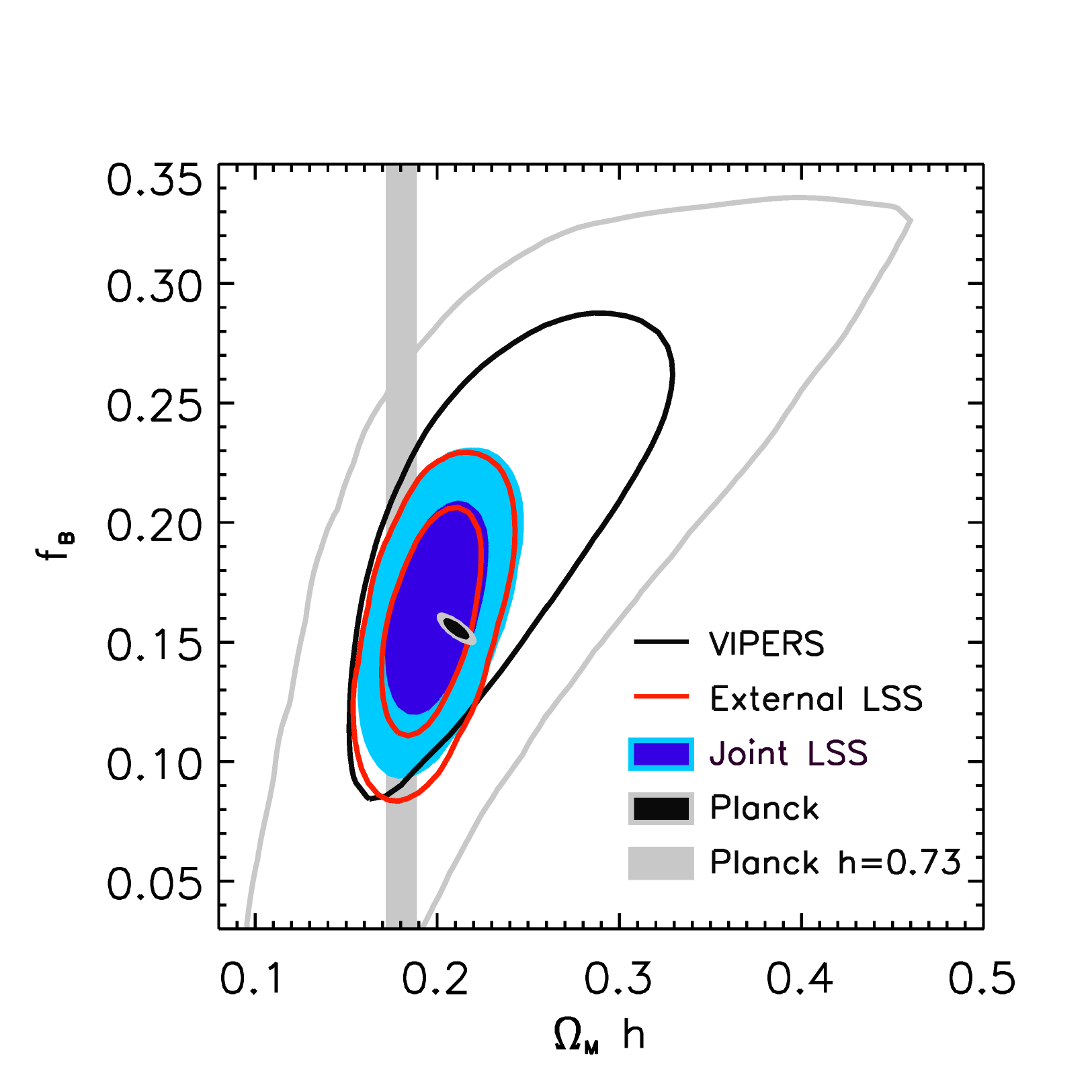}
\end{center}
\caption{Joint parameter constraints from LSS surveys including 2dFGRS, SDSS LRG, WiggleZ, and VIPERS.  The combined constraint including VIPERS is indicated by the solid contour.  The reference from Planck is indicated by the ellipse.  Rescaling the constraint on $\Omega_M h^3$ from the Planck temperature power spectrum using the local estimate of $H_0$ gives a prior on $\Omega_M h$ indicated by the vertical grey band.
  \label{fig:joint}}
\end{figure}

We compute the combined constraints from the external LSS surveys consisting of 2dFGRS, SDSS LRG and WiggleZ as shown in Fig. \ref{fig:joint}.  We find this constraint to be fully consistent with VIPERS.  Combining with the VIPERS likelihood gives the best available constraints from the LSS surveys to redshift $z=1.1$.  We find marginalised values: $\Omega_M h = 0.206_{-0.015}^{+0.013}$ and $f_B = 0.170 _{-0.025}^{+0.028}$.
These values are consistent with the Planck ones, $\Omega_M h = 0.211\pm 0.004$
and $f_B = 0.158 \pm 0.002$ \citep{planck2015}
within the statistical uncertainties.

It is interesting to ask if our determination of $\Omega_M h$ can shed any light on the disagreement concerning $H_0$ between Planck and direct measurements.
For flat cosmological models, the angular location of the acoustic scale in the CMB temperature power spectrum is approximately sensitive to the parameter combination $\Omega_M h^3$, and this quantity should be robust even in the face of small scale-dependent systematics in Planck, which have been proposed as a possible explanation for the $H_0$ tension \citep{addison2016}.
Using the Planck temperature likelihood routine in CosmoMC we determine the marginalised value $\Omega_M h^3=0.0965  \pm  0.0005$. Here we assume that the peak location is the dominant source for this constraint. Adopting the local estimate $H_0=73.24 \pm  1.74\ \rm{km\,s^{-1}\,Mpc^{-1}}$ \citep{Riess2016} leads to a low value of $\Omega_M h=0.180\pm0.009$ indicated by the vertical shaded band in Fig. \ref{fig:joint}. This value is displaced from the full Planck constraint due to the 3.5$\sigma$ tension in the best-fit value of $H_0$.  The combined LSS constraints lie between Planck and this lower value, and are consistent with both. The precision of current data therefore does not permit LSS to adjudicate in the $H_0$ dispute -- but
this diagnostic will sharpen with data from future surveys of larger volumes, and this is one way in which the $H_0$ debate could be resolved.

%
%FIGURE 

\subsection{Summary and Conclusions}

In this paper we have presented the first measurement of the galaxy power spectrum from a sample extending beyond redshift $z=1$, using the final data from the VIMOS Public Extragalactic Redshift Survey (VIPERS). 
In particular

\renewcommand{\labelitemi}{$\bullet$}
\begin{itemize}

\item we have discussed and tested in detail how the geometry and selection function of the VIPERS survey can be modelled, yielding an accurate description of the corresponding window function in Fourier space; 

\item we have tested and validated the corrections for all observation-specific effects affecting the VIPERS data, using a large set of custom-built mock samples. We similarly assessed the degree of modelling uncertainties related to non-linear clustering, galaxy biasing and redshift-space distortions. We show that residual systematic errors on the cosmological parameters deriving from these effects are about 20 times smaller than the statistical errors; 

\item we have presented new measurements of the power spectrum of galaxy clustering using 51,728 galaxies distributed within four independent subsamples defined by two redshift ranges $0.6<z<0.9$ and $0.9<z<1.2$ over the two VIPERS fields W1 and W4;

\item we have used the set of mocks to estimate covariance matrices for all the measurements, and to access the range of scales where the effects of non-linear evolution on the shape of the power spectrum can be considered to be under control;

\item we have used these ingredients to fit the data with a cosmological model for $P(k)$ with three free cosmological parameters ($\Omega_M,f_B,h$) and three parameters that encode galaxy physics (bias in each redshift bin and velocity dispersion); combining the four power spectrum measurements, this yields an estimate of the mean value of the matter density (scaled to the current epoch), $\Omega_M h=0.227^{+0.063}_{-0.050}$, and  baryon fraction $f_B=0.220^{+0.058}_{-0.072}$, after marginalising over galaxy bias;

\item these values, which describe the galaxy distribution when the Universe was about half its current age, are in agreement with measurements at lower redshift from 2dFGRS at $z=0.2$, SDSS LRG at $z=0.35$, and WiggleZ at $0.2 < z < 0.8$. We further demonstrate consistency with the Planck determination of $\Omega_M h$ and $f_B$;

\item comparison to previous configuration space constraints on $\Omega_M$ from VIPERS (counts in cells) shows consistency despite the intrinsically different nature of the measurements and their covariances.

\end{itemize}

These results have extended the classical cosmological test of determining the matter content of the Universe from the shape of the galaxy power spectrum. There is no reason to believe that this method has reached the limit of its precision, and we expect the error contours to continue to shrink with new generations of larger galaxy surveys. In this way, the galaxy power spectrum has the potential to clarify current areas of cosmological uncertainty, such as the true value of $H_0$.